\documentclass[twocolumn]{aastex61}
\usepackage{color}
\received{}
\revised{}
\accepted{}
\submitjournal{ApJS}

\shorttitle{Neutrino transport with MC method}
\shortauthors{Kato et al.}
\begin{document}

%\title{Implementation of nucleon recoils in core-collapse supernovae simulations}
\title{Neutrino transport with Monte Carlo method: I. Towards fully consistent implementation of nucleon recoils in core-collapse supernova simulations}

\correspondingauthor{Chinami Kato}
\email{chinami.kato.e8@tohoku.ac.jp}

\author{Chinami Kato}
\affiliation{Department of Aerospace Engineering, Tohoku University, 6-6-01 Aramaki-Aza-Aoba, Aoba-ku, Sendai 980-8579, Japan}

\author{Hiroki Nagakura}
\affiliation{Department of Astrophysical Sciences, Princeton University, Princeton, NJ 08544}

\author{Yusuke Hori}
\affiliation{School of Advanced Science and Engineering, Waseda University, 3-4-1, Okubo, Shinjuku, Tokyo 169-8555, Japan}

\author{Shoichi Yamada}
\affiliation{School of Advanced Science and Engineering, Waseda University, 3-4-1, Okubo, Shinjuku, Tokyo 169-8555, Japan}
\affiliation{Advanced Research Institute for Science and Engineering, Waseda University, 3-4-1, Okubo, Shinjuku, Tokyo 169-8555, Japan}

\begin{abstract}
The small energy exchange via nucleon recoils in neutrino-nucleon scattering is now supposed to be one of the important factors for successful explosion of core-collapse supernovae (CCSNe) as they can change neutrino spectra through accumulation of a large number of scatterings. %, although the energy exchange in an individual scattering is small.
%It is known that they can change neutrino spectra through accumulation of a large number of scatterings%, although the energy exchange in an individual scattering is small.
In finite-difference methods employed for neutrino transport in CCSN simulations, we normally can not afford to deploy a large enough number of energy bins needed to resolve this small energy exchange and sub-grid techniques are employed one way or another. 
In this paper we study quantitatively with the Monte Carlo (MC) method how well such a treatment performs.
We first investigate the effects of nucleon recoils on the neutrino spectra and confirm that the average energy is reduced by $\sim$15\% for heavy-lepton neutrinos and by much smaller quantities for other types of neutrinos in a typical post-bounce situation.
It is also observed that the nucleon scattering dominates the electron scattering in the thermalization of neutrino spectra in all flavors. 
We then study possible artifacts that the coarse energy grid may produce in the finite-difference methods.
In order to mimic the latter calculation, we re-distribute MC particles in each energy bin after a certain interval in a couple of ways and study how the results are affected and depend on the energy-resolution.
We also discuss possible implications of our results for the finite-difference methods.

%We introduce an energy grid with 10 and 20 meshs in the Monte Carlo simulation, which is mimic to the Boltzmann solver, and impose four artificail distributions to energy spectra. 
%We find that the piece-wise linear distributions, ensureing the number conservation of neutrinos in each energy bin, the difference from the orifinal ones is within a few \% in the case of 20 meshes. 
%If we reduce the number of grids to 10, the energy conservation becomes important. 
%The difference in the case satisfing both conservations is still $\sim$10\%, whereas spectra without the energy conservation deviate by 20\%. 
%We find that the energy conservation in each energy bin is necessary condition for the precise treatments of nucleon recoils in the finite-difference neutrino transport with low-energy resolution.
\end{abstract}

%and average the energy distribution in each energy bin to impose artificially piece-wise constant energy spectra in the Monte Carlo calculations.
%We find that the obtained spectra are deviated from the original ones by $\lesssim$ 20\% owing to the overestimation of energy exchange.
%If we introduce piece-wise linear distributions instead, ensuring the conservations of number and energy of neutrinos in each energy bin, the difference is reduced to a few \%.

%The spectra of anti-electron neutrinos are not changed by nucleon recoils owing to the small energy exchange, although the number of nucleon scattering is larger than those of other neutrino reactions.



%\keywords{editorials, notices --- 
%miscellaneous --- catalogs --- surveys}

%% That way, if you change the order of any elemesnts, LaTeX will


\section{Introduction}

%%about SN
Core-collapse supernovae (CCSNe) are violent explosions of massive stars with $M_{\mathrm{ZAMS}} \gtrsim 8\ M_\odot$.
The explosion is instigated by the gravitational collapse of a central core, which is followed by the formation of a shock wave at core bounce.
If the shock wave passes through the central core and propagates through outer envelopes up to the stellar surface, these envelopes are ejected and a compact remnant is left behind at the center.
In numerical simulations, the shock wave stagnates inside the core and how to get the shock wave out of the core has been explored for a long time but has not been settled yet \citep[references therein]{2012ARNPS..62..407J,2012AdAst2012E..39K,2019arXiv190411067M}.
One of the favored mechanisms for shock revival is the heating by neutrinos emitted from a proto-neutron star (PNS) and is called the neutrino heating mechanism.
In multi-dimensional simulations, non-spherical matter motions, such as convection or the standing accretion shock instability (``SASI''), push up the shock wave and enhance the neutrino heating behind it \citep{2003ApJ...584..971B,2008ApJ...678.1207I}, and shock revival is obtained more often than not recently \citep{Skinner:2015uhw,Summa:2015nyk,2015ApJ...801L..24M,2015ApJ...807L..31L,2015ApJ...800...10D,Takiwaki:2016qgc,2016ApJ...831...98R,2017MNRAS.472..491M,2017ApJ...850...43R,OConnor:2015rwy,2018ApJ...855L...3O,2019MNRAS.482..351V,2019MNRAS.485.3153B,adam2020}.

%%importance of improvement of neutrino opacity
%%importance of nucleon recoils
Neutrino reaction rates are certainly important for SN explosion.
\cite{1985ApJS...58..771B} provided a comprehensive set of neutrino opacities, which have been widely incorporated in SN simulations.
Possible corrections to these rates have been investigated for the last 30 years.
For example, the important updates are summarized in \cite{2018ApJ...853..170K} (see also references therein).
They have been taken into account in numerical simulations of late \citep{2006A&A...447.1049B,2012ApJ...761...72M,2012ApJ...760...94L,2018ApJ...853..170K}.

Nucleon recoils in neutrino-nucleon scattering are one of them.
Since the energy exchange by nucleon recoils is only a few \% of initial neutrino energy owing to the nucleon mass much larger than the typical neutrino energy $\lesssim$~100 MeV, they were considered to be less important in the spectral formation than electron scattering, in which the energy exchange is much more efficient, and ignored in the past SN simulations.
The cross section of nucleon scattering is much larger than that of electron scattering, however, and it is possible that neutrino spectra are changed by nucleon recoils, especially for heavy-lepton neutrinos, which interact with matter only via neutral current reactions.
%Electron and anti-electron neutrinos experience many times of charged current reactions and they contribute to the thermalization of spectra, mainly.
% and neutrino spectra deviate from the thermal equilibritum at the deeper position than electron type neutrinos.
%There is a diffusion region, in which neutrino spectra are changed by nucleon scatterings (``scattering atmosphere'') \citep{Raffelt:2001kv}.
As a matter of fact, the effects of nucleon recoils have been already investigated.
For example, \cite{Keil:2002in} used their Monte Carlo (MC) code for the assessment and demonstrated that the average neutrino energy is indeed decreased by nucleon recoils. 
Their effects have been also studied by dynamical simulations of CCSNe \citep{2002A&A...396..361R,2006A&A...447.1049B,2009ApJ...694..664M,2010PhRvL.104y1101H,2012ApJ...760...94L,2012ApJ...761...72M,2015ApJ...808..188P,2015ApJ...807L..31L,Skinner:2015uhw,2017ApJ...850...43R,2018ApJ...853..170K,2018arXiv180905608B,2019MNRAS.482..351V,2019MNRAS.485.3153B,2019arXiv190110523R,2019ApJ...873...45G}.
They found that nucleon recoils reduce the opacity for neutrinos and accelerate the PNS cooling, which in turn increases neutrino luminosities, thus helping shock revival.

%%difficulty to take into account
We revisit this issue from a bit different point of view.
In most of CCSNe simulations one employs a finite-difference method for neutrino transport.
In so doing, we normally can not afford to deploy a sufficiently large number of energy bins needed to resolve the small energy exchange by nucleon recoils.
For example, only 20 energy bins are deployed to cover the range of 0-300 MeV in our CCSN simulations with full Boltzmann neutrino transport \citep{nagakura2018,2019ApJ...880L..28N,harada2019} and the widths of these energy bins are larger by an order than the typical energy exchange through nucleon recoils.
Note that although in those simulations energy sub-grids are normally employed to evaluate the transfer rate from an energy cell to the next one \citep{2006A&A...447.1049B}, the resolution problem still remains, since the neutrino distribution in the energy bin is not assumed one way or another.
We will quantify the effects of the coarse energy grid and present a possible improvement in this paper.

We perform neutrino transport calculations with our own MC code for a static hydrodynamical background derived from our dynamical SN simulation.
Note that these MC simulations are free of the energy-resolution problem.
It is also mentioned that in this study we do not use the approximation given by \cite{Horowitz:2001xf} but employ the exact reaction rate for nucleon scattering\footnote{Note that we neglect the effect of weak magnetism, which is embedded in the form factor of the scattering kernel, in order to purely focus on the effects of nucleon recoils in this study. The incorporation of the weak magnetism in our MC code is straightforward, though.}.
After validating our MC code, we look into the effects of nucleon recoils on neutrino spectra, that is, how they are thermalized with radius, comparing their contributions with others, particularly electron scattering, in detail.
%In this paper, we show the radial profile of neutrino distribution functions and pay attention to the comparison between nucleon- and electron scatterings, especially.
We then assess the energy-resolution issue by introducing energy grids with different numbers of grid points: $N_{E_\nu}$ = 10 and 20 in our MC calculations to assess the energy-resolution issue.
Note that the latter energy grid is exactly the same as the one used in our CCSN simulations with the finite-difference Boltzmann solver.
In order to mimic the situation in the finite-difference methods, we re-distribute by hand in a couple of ways the MC particles in each energy bin repeatedly after some periods given by the typical time step of CCSN simulations and see their effects on neutrino spectra.

%We consider four artificial models of the neutrino distribution: flat, linear, linear+Ncons and linear+NEcons (see Section~\ref{ch5}) and redistribute sample particles in each energy bin following these distributions repeatedly on the time-steps of CCSN simulations to mimic the situation in the finite-difference method and see its effects on neutrino spectra.
%We artificially homogenize the neutrino distribution in each energy bin repeatedly on the time-steps of CCSN simulations to mimic the situation in the finite-difference method and see its effects on neutrino spectra.
%Finally, instead of simply homogenizing the neutrino distribution in the bin, we consider a linear distribution that ensures number and energy conservations in the energy bins and evaluate quantitatively if to what extent the numerical error is reduced.

%%outline
The organization of the paper is as follows: the new features in our MC code are briefly described in Section \ref{ch2}, particularly the treatment of neutrino-nucleon scattering; several numerical tests for the validation of our new code are presented in Section \ref{ch3}; the effects of nucleon recoils on neutrino spectra are discussed in Section \ref{ch4}; the possible influence of energy resolution in the finite-difference methods is studied in Section \ref{ch5}, and finally we give summary and discussions in Section \ref{ch6}.

%%%%%%%%%%%%%%%%%%%%%%%%%%%%%%%%%%%%%%%%%%%%%%%%%%%%%%%%%%%%%%%%%%%%%%%%%%%%%%%%%%%%%%%%%%%%%%%%%%%%%%%%%%%%%%%%%%%%%%%%%
%%%%%%%%%%%%%%%%%%%%%%%%%%%%%%%%%%%%%%%%%%%%%%%%%%%%%%%%%%%%%%%%%%%%%%%%%%%%%%%%%%%%%%%%%%%%%%%%%%%%%%%%%%%%%%%%%%%%%%%%%%

%%%%

\begin{table*}[htbp]
\caption{The neutrino reaction set included in our calculations. The base model incorporate the sub-set of neutrino reactions normally considered in dynamical supernova simulations. 
The nucleon recoil in the nucleon scattering is taken into account in model r1 whereas the electron/positron scattering is also included in model e1.
\label{reac_MC}}
\begin{center}
\begin{tabular}{l|l|l|ccc} \hline
\multicolumn{3}{c|}{reactions}                                                                    & base & r1 & e1 \\
\hline\hline
electron-positron pair annihilation & pair      & $e^- + e^+ \longrightarrow \nu + \bar{\nu} $    & $\checkmark$ & $\checkmark$ &$\checkmark$\\
bremsstrahlung                      & brems     & $N + N \longrightarrow N + N + \nu + \bar{\nu}$   & $\checkmark$ & $\checkmark$ &$\checkmark$\\
electron capture                    & ecp       & $p + e^- \longleftrightarrow n + \nu_e$         & $\checkmark$ & $\checkmark$ &$\checkmark$\\
positron capture                    & pc        & $n + e^+ \longleftrightarrow p + \bar{\nu}_e$   & $\checkmark$ & $\checkmark$ &$\checkmark$\\
\hline
nucleon scattering                  & nsc (Bruenn)  & $N + \nu \longrightarrow N + \nu$           & $\checkmark$ &         & \\
                                    & nsc (rec)  &                                                &        & $\checkmark$ &$\checkmark$\\
electron scattering                 & esc       & $e^- + \nu \longrightarrow e^- + \nu$            &        &        &$\checkmark$\\
positron scattering                 & psc       & $e^+ + \nu \longrightarrow e^+ + \nu$            &        &       &$\checkmark$\\
\end{tabular}
\end{center}
\end{table*}

%%%%

\section{Numerical methods of MC transport} \label{ch2}
\subsection{MC method $\text{vs}$ finite-difference methods}
There are two \textcolor{black}{representative} approaches to the numerical solution of the radiation transport equation: the discretized methods and the MC method.
In the former method, such as the  $S_N$ method (see e.g. \cite{2004rahy.book.....C}), we discretize the transport equation in phase space.
In the latter method, we follow the tracks of ``sample particles'', which represent a bundle of radiation particles interacting with matter.
The interactions are treated probabilistically and physical quantities, such as the distribution function of radiation, are obtained by collecting individual sample evolutions.
Each method has its own advantages and drawbacks.

In the discretized method, it is normally no problem to treat the entire system having both optically thick and thin regions. 
The time-dependent coupling with hydrodynamics is also straightforward. 
On the other hand, the numerical resolution is mainly determined by the number of mesh points one can afford and, as repeatedly mentioned, the energy-grid number cannot be very large particularly in multi-spatial-dimensions. 
This may be particularly critical for the treatment of the small energy exchanges in the nucleon scattering and special cares, such as the employment of sub-grids, are taken normally \citep{2006A&A...447.1049B,2018arXiv180905608B}.
Recently, \cite{2019arXiv190405047S} shows that the Fokker-Planck approximation is also useful. 
It is noted that even if such a measure is taken, the coarse-resolution problem may remain, since the neutrino energy spectrum is still represented on the rather small number of energy-grid points.

The MC method is mesh-free and hence favorable for multi-dimensional simulations.
Various reactions can be treated in a simple and direct way.
In fact, the small energy exchanges in the nucleon scattering pose no problem in this approach.
On the other hand, statistical errors inherent to the probabilistic description and slow convergence scaled as $\sqrt{N}$ are big disadvantages for the MC method.
It is normally counted as another demerit that it is difficult to treat optically thick regime and/or couplings with hydrodynamics (but see \cite{2012ApJ...755..111A,2017ApJ...847..133R}).

In this study, we employ the MC method for neutrino transport for two reasons.
First, we focus on nucleon recoils, which can be treated most accurately with the MC method as explained above. %suitable for MC methods.
Second, we are concerned with the thermalization of neutrino spectrum via the nucleon scattering and hence we do not need to worry about the high density region, where the MC method performs poorly.
As a matter of fact, neutrinos are already thermalized by other processes well inside the neutrino sphere and we have only to impose the thermal distribution functions as the inner boundary condition (but see Section~\ref{subch:thermal_neutrino} for more details of our treatment).
%Due to the density and temperature decreases, the number of reactions also decreases and neutrinos become free-streaming at the outer part of the core.
%Nucleon scatterings may hence affect neutrino spectra between two regions and we do not need to solve the wide region, on which is usually focused in dynamical SN simulations.

%The MC approach is usually applied to solving the transport of photons \citep{1999A&A...344..282L,1999A&A...345..211L,Kasen:2006ce,Maeda:2005pi}.
%The basic ideas for neutrinos are same as those for photons, except for the Fermi-blocking of neutrinos.
%For example, the emission rates of neutrinos depend on the blocking factor $1-f$ with the distribution function $f$.
%It changes with time due to absorption, emission or scattering by matters and we have to update it within an appropriate timescale (See Sections~\ref{subch:trans}, \ref{subch:f}).
%The MC method adopted to the neutrino transport in CCSNe by several numerical studies. 
%\cite{1978ApJS...37..287T} investigated the thermalization of neutrino spectra in static uniform matter.
%The steady-state calculations on 1D SN hydrodynamical backgounds were done by \cite{1989A&AS...78..375J}.
%Spectral formation of neutrinos was discussed by \cite{Keil:2002in}.
%\cite{2012ApJ...755..111A} extended the implicit MC method to the optically thick region and coupled to hydrodynamical euqations.
%Some numerical studies used the MC method for the evaluation of the discretized method, with Boltzmann solver \citep{1999A&A...344..533Y,2017ApJ...847..133R} and with M1 closure scheme \citep{2017MNRAS.469.1725M}.

\subsection{New features in our MC code} \label{new_MC}

Here we summarize some new features of our MC code worth particular mention.
Other information on the code is provided in Appendices \ref{appendix}-\ref{appendix3}.
%We have developed a new MC code for spherical symmetric neutrino transport on a static background in order to calculate the exact neutrino spectra.
%In this section, we briefly describe numerical treatments, especially their new features, adopted in our code.

The basics are essentially the same as in previous works \citep{1978ApJS...37..287T,1989A&AS...78..375J,Keil:2002in}.
The main difference in the neutrino transport from the photon transport is the Fermi-blocking at the final state. 
For example, neutrino scatterings are suppressed by the blocking factor $1-f$, where the distribution function is denoted by $f$. 
This makes the transport equation nonlinear and we need to update the distribution function at an appropriate rate during the MC simulation (see Appendices \ref{subch:trans} and \ref{subch:f}).

In our code, four emission and two scattering processes are implemented (see Table~\ref{reac_MC}).
Here we focus on the nucleon scattering, the key reaction in this paper.
As mentioned earlier, we treat this process as precisely as possible. 
We do not use the approximate formula commonly used but employ the exact reaction rate, which is essentially the same as for the electron scattering. 
We store it in a table as $R_i(E_\nu, E^\prime_\nu, \psi)$ for various combinations of density, temperature and electron fraction.
In this expression, $E_\nu$ and $E^\prime_\nu$ are the neutrino energies before and after scattering, respectively; $\psi$ is the scattering angle, i.e., the angle that the incident and outgoing momenta make. 
The table actually contains the reaction rates only for $E_\nu \le E^\prime_\nu$ and the other case $E_\nu > E^\prime_\nu$ is derived from the former so that the detailed balance relation should be satisfied.
The detailed procedure is given in Appendix \ref{appendix}.

For a given incident energy $E_\nu$, the scattering angle $\psi(\theta^\prime_\nu, \phi^\prime_\nu)$ and the energy after scattering $E^\prime_\nu$ are determined probabilistically according to their normalized distributions $P_\psi$ and $P_{E^\prime_\nu}$, which are derived from the cumulative reaction rate $R_i(E_\nu,E^\prime_\nu,\psi)$ (see eqs. (\ref{Ppsi}) and (\ref{penu}) ).
The azimuth of the scattering direction $\Psi$ is determined randomly in the range of [0, 2$\pi$]. 
Then, the propagation direction of neutrinos after scattering in phase space specified by the zenith and azimuth angles measured from the local radial direction, ($\theta^\prime_\nu, \phi^\prime_\nu$), is given from the angles ($\psi, \Psi$) by an appropriate coordinate transformation.

Note that the normalized distributions $P_\psi$ and $P_{E^\prime_\nu}$ do not include the blocking factor $1-f$ (see Section~\ref{NNscat}).
It is taken into account after $E^\prime_\nu$, $\theta^\prime_\nu$ and $\phi^\prime_\nu$ are determined in this way. 
We throw a dice yet again to get a random number $z$ in the range of [0,~1].
If the condition $0 \leq z \leq f(r,E^\prime_\nu,\theta^\prime_\nu)$ is satisfied, we accept this scattering whereas it is "blocked'' otherwise and the energy and angles of neutrinos are not changed after all.
Note that this procedure correctly reproduces the mean free path in the presence of Fermi-blocking. 
It has an advantage that the reaction table can be independent of the neutrino distribution.

%Although these mean free paths of sample particles seem to be longer than the correct ones, it does not matter if we derive the distance to the next neutrino reaction (the reaction lengths $l_{\rm{r}}$) from the cross sections $\sigma_i$ without the blocking factor.
%In this case, the reaction lengths are estimated shorter than the correct mean free paths and sample particles interact with matters at the correct frequency, consequently.
%In other words, we divide the cross section into two parts: blocking factor and the other part.  
%The main reason why we adopt this complicated method for scattering processes is that we have to include their reaction rates as a table in our code.

\subsection{reaction rate of neutrino-nucleon scattering} \label{NNscat}

The reaction rate of the neutrino-nucleon scattering is given essentially in the same way as for the electron scattering \citep{1993ApJ...410..740M}:
\begin{eqnarray}
R_{\rm{rec}}\left(q,q^\prime\right) = \frac{G^2_F}{2\pi^2\hbar c} \frac{1}{E_\nu E_{\nu}^\prime}
                          \left[\beta_1I_1 + \beta_2 I_2 + \beta_3I_3\right]. \label{R_rec}
\end{eqnarray}
In the above expression, $G_F = 1.166364 \times 10^{-11} \rm{MeV}^{-2}$ is the Fermi coupling constant and $\beta$'s are the following combinations of the coupling constants:
 $\beta_1 = \left(C_V - C_A \right)^2$, $\beta_2 = \left(C_V + C_A \right)^2$ and $\beta_3 = C^2_A - C^2_V$,
 and $I$'s are functions of the energies $E_\nu$, $E^\prime_\nu$ of the incident and outgoing neutrinos and the angle $\psi$ between their momenta $q$ and $q^\prime$:
\begin{eqnarray}
I_1 &=& \frac{2\pi T}{\Delta^5} E^2_\nu E_{\nu}^{\prime2} (1-\cos{\psi})^2 
      \frac{1}{\exp{\left(\frac{E_\nu - E_\nu^\prime}{T}\right)}-1} \nonumber \\
&&      \times \left[ AT^2\left(G_2(y_0) + 2y_0G_1(y_0) + y_0^2G_0(y_0)\right)\right. \nonumber \\
&&      \left. + BT\left(G_1(y_0) + y_0G_0(y_0) \right) + CG_0(y_0) \right], \\
I_2 &=& I_1\left(-q,-q^\prime \right), \\
I_3 &=& \frac{2\pi T m_N^2}{\Delta}E_\nu E_\nu^\prime\left(1-\cos{\psi}\right) 
      \frac{G_0\left(y_0\right)}{\exp{\left(\frac{E_\nu - E_\nu^\prime}{T}\right)}-1},
\end{eqnarray}
with 
\begin{eqnarray}
\Delta^2 &\equiv& E_\nu^2 + E_\nu^{\prime^2}-2E_\nu E_\nu^\prime\cos{\psi}, \\
A &\equiv& E_\nu^2 + E_\nu^{\prime2} + E_\nu E_\nu^\prime\left(3+\cos{\psi}\right), \\
B &\equiv& E_\nu^\prime \left[ 2E_\nu^{\prime2} + E_\nu E_\nu^\prime\left(3-\cos{\psi}\right) \right. \nonumber \\
&&\ \ \   \left.  - E_\nu^2\left(1+3\cos{\psi}\right)\right], \\
C &\equiv& E_\nu^{\prime2} \left[ \left(E_\nu^\prime-E_\nu\cos{\psi}\right)^2 - \frac{E_\nu^2}{2}\left(1-\cos^2{\psi}\right) \right. \nonumber \\
  &&\ \ \  \left.  - \frac{1}{2}\frac{1+\cos{\psi}}{1-\cos{\psi}}\frac{m_N^2}{E_\nu^{\prime2}}\Delta^2\right], \label{R_rec_fin}
\end{eqnarray}
and $y_0 = E_{N0}/T$, $\eta=\mu_N/T$, $\eta^\prime=\eta+(E_\nu-E^\prime_\nu)/T$ and $G_n(y) \equiv F_n(\eta^\prime-y) - F_n(\eta-y)$, in which the Fermi integral $F_n(z)$ is defined as
\begin{eqnarray}
F_n\left(z\right)=\int_0^\infty \frac{x^n}{e^{x-z}+1}dx ,
\end{eqnarray}
and $E_{N0}$ is expressed as 
\begin{eqnarray}
E_{N0} = \frac{E_\nu-E_\nu^\prime}{2} + \frac{\Delta}{2}\sqrt{1+\frac{2m_N^2}{E_\nu E^\prime_\nu\left(1-\cos{\psi}\right)}}. \label{R_rec_f}
\end{eqnarray}

\textcolor{black}{Assuming that} the energy exchange is much smaller than the neutrino energy before scattering $\Delta E/E_\nu \ll 1$ and the nucleon mass is infinitely large $m_N \rightarrow \infty$, one reproduces the reaction rate given by \cite{1985ApJS...58..771B}, which is commonly incorporated in SN simulations:
\begin{eqnarray}
  &&\textcolor{black}{R_{\rm{Bruenn}}} = \frac{2\pi G^2_F}{\hbar c} \eta_{NN} \delta\left(E_\nu-E^\prime_\nu\right) \nonumber \\
  &&\times \left\{ \left(h^N_V\right)^2 + 3\left(h^N_A\right)^2 + \left[\left(h^N_V\right)^2 - \left(h^N_A\right)^2 \right] \cos{\psi} \right\}, \ \ \ \label{bruenrate}
\end{eqnarray}
and $\eta_{NN}$ is defined as 
\begin{eqnarray}
\eta_{NN} &\equiv& \int \frac{2d^3 p_N}{\left(2 \pi\right)^3} \tilde{F}_N\left(\tilde{E}\right) \left[1-\tilde{F}_N\left(\tilde{E}\right)\right],
\end{eqnarray}
where $\tilde{F}_N(\tilde{E}) = 1/[1+\exp{(\tilde{E}-\mu_N)/T}]$ is the Fermi-Dirac distribution of nucleons with the non-relativistic energy $\tilde{E}=p^2_N/2m_N$.

The exact and (Bruenn's) approximate total cross sections are obtained by integrating the corresponding reaction rates $R_\ast = R_{\rm{rec}}, \textcolor{black}{R_{\rm{Bruenn}}}$:
\begin{eqnarray}
  \sigma_{\rm{N}} = \int \tilde{R}_\ast d\cos{\psi},
\end{eqnarray}
with
\begin{eqnarray}
\tilde{R}_\ast = \frac{1}{\left(2\pi\right)^3} \int 2\pi E^{\prime2}_\nu R_\ast dE^\prime_\nu.
\end{eqnarray}

The quantities after scattering $E^\prime_\nu$, $\cos{\theta^\prime_\nu}$ and $\phi^\prime_\nu$ are determined as follows.
We first determine the scattering angle $\psi$ according to the normalized cumulative distribution:
\begin{eqnarray}
  &&P_\psi\left(\cos{\psi_k}; E_\nu\right) \nonumber \\
  && = \frac{\int_{-1}^{\cos{\psi_k}}\int 2\pi E^{\prime2}_\nu R_\ast\left(E_\nu, E^\prime_\nu,\cos{\psi}\right) dE^\prime_\nu d\cos{\psi} }{\int^1_{-1}\int 2\pi E^{\prime2}_\nu R_\ast\left(E_\nu,E^\prime_\nu,\cos{\psi}\right) dE^\prime_\nu d\cos{\psi}}.\ \ \ \ \label{Ppsi}
\end{eqnarray}
For the derived $\psi_k$, the energy after scattering is determined in the same way according to the following normalized cumulative distribution:
\begin{eqnarray}
&& P_{E^\prime_\nu} \left(E^\prime_{\nu,i}; \cos{\psi_k}, E_\nu\right) \nonumber \\
&& \ \ \ \  = \frac{\int_{E^\prime_{\rm{min}}}^{E^\prime_{\nu,i}} 2\pi E^{\prime2}_\nu R_\ast\left(E_\nu,E^\prime_{\nu},\cos{\psi_k}\right) dE_\nu^\prime}{\int_{E^\prime_{\rm{min}}}^{E^\prime_{\rm{max}}} 2\pi E^{\prime2}_\nu R_\ast\left(E_\nu, E^\prime_\nu,\cos{\psi_k}\right) dE^\prime_\nu}. \label{penu}
\end{eqnarray}
The minimum and maximum energies $E^\prime_{\rm{min}}$, $E^\prime_{\rm{max}}$ in the integration are determined so that the reaction rates there should be $10^{-5}$ times smaller than the maximum rate.

The treatments of other reactions are summarized in Appendices~\ref{appendix2} and \ref{appendix3}.

\begin{figure}[htbp]
\center
\epsscale{1.4}
\plotone{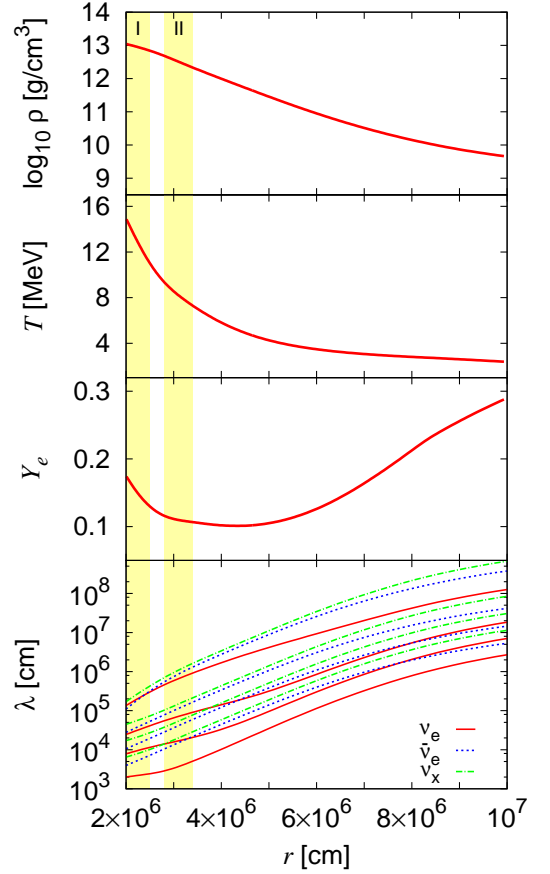}
\caption{The radial profiles of density, temperature, electron fraction and total mean free paths for different species of neutrinos in the progenitor model with $M_{\rm{ZAMS}}$ = 11.2~$M_\odot$ at 100 ms after core bounce \citep{2019ApJS..240...38N}. The mean free paths for each species are shown for $E_\nu$ = 5, 14, 24 and 40 MeV (from above) with the same color for the r1 set of neutrino reactions (see Table~\ref{reac_MC}). We focus on the regions painted in yellow in the comparison. \label{hydro}}
\end{figure}

\begin{figure*}[htbp]
  \epsscale{0.9}
  \plottwo{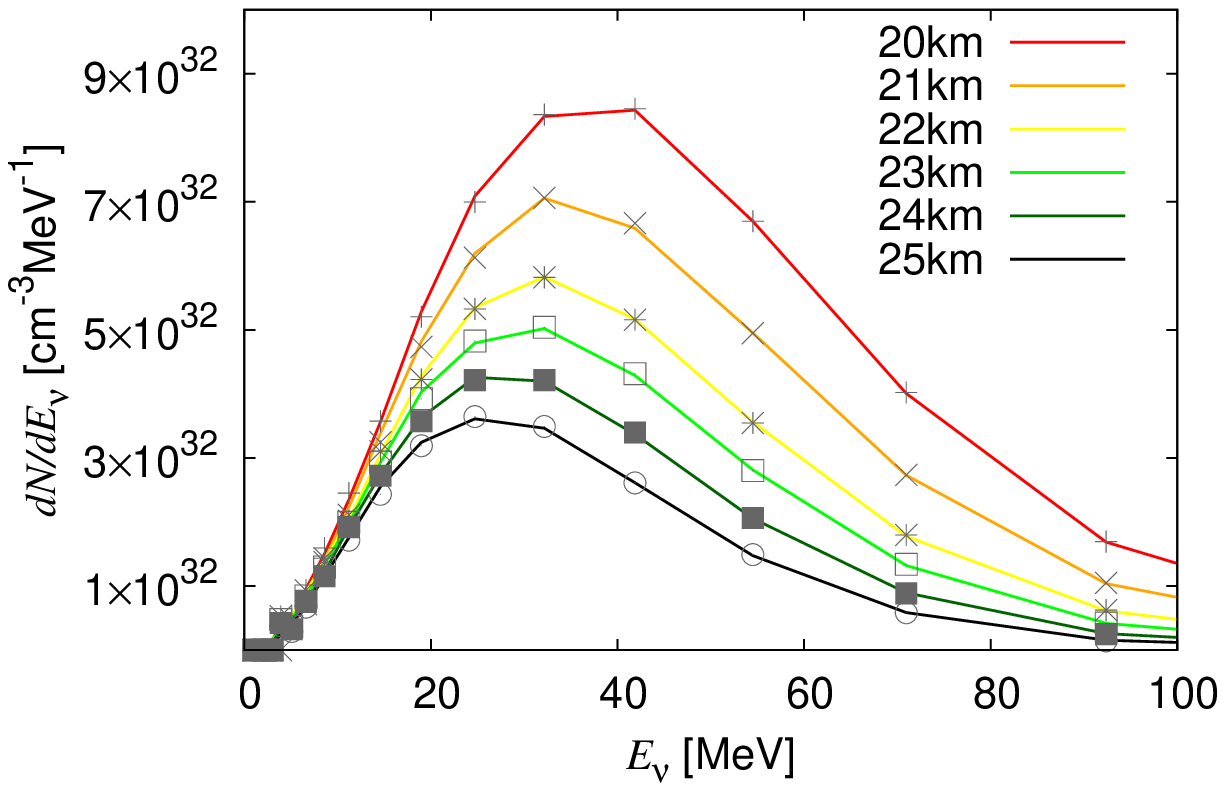}{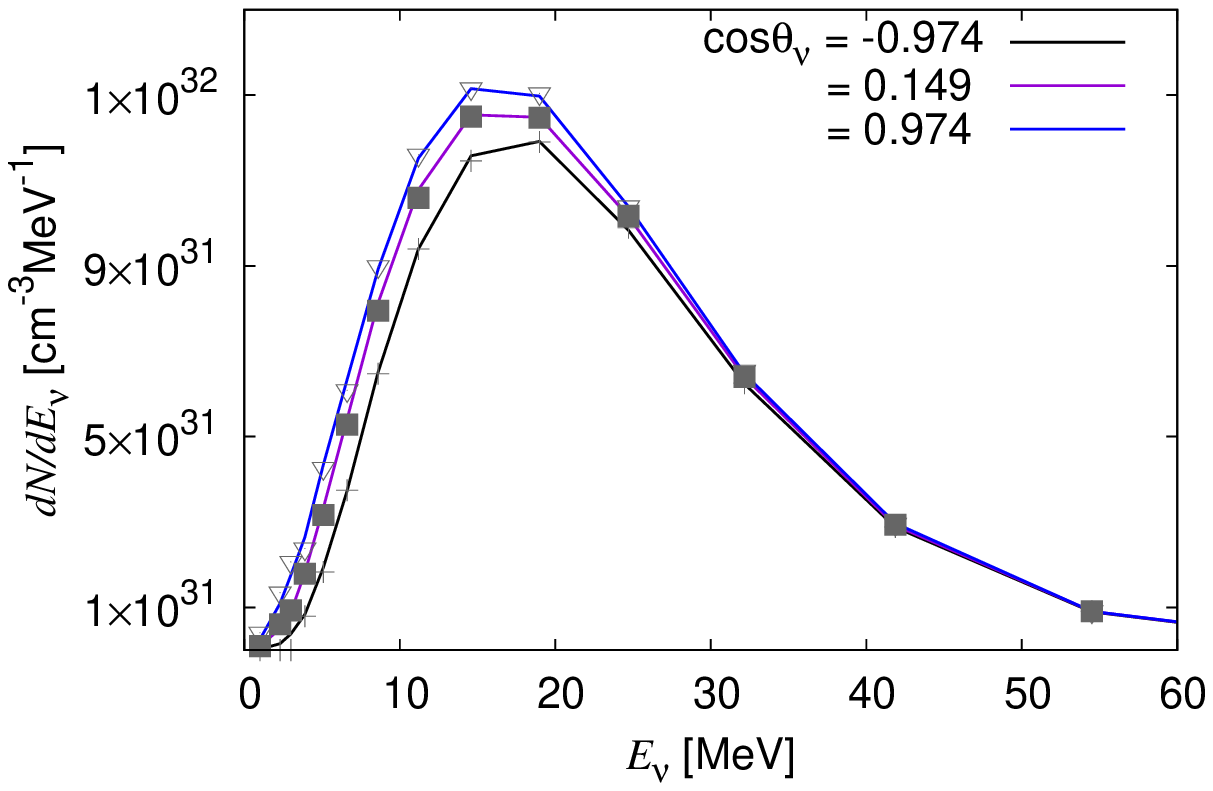}
  \plottwo{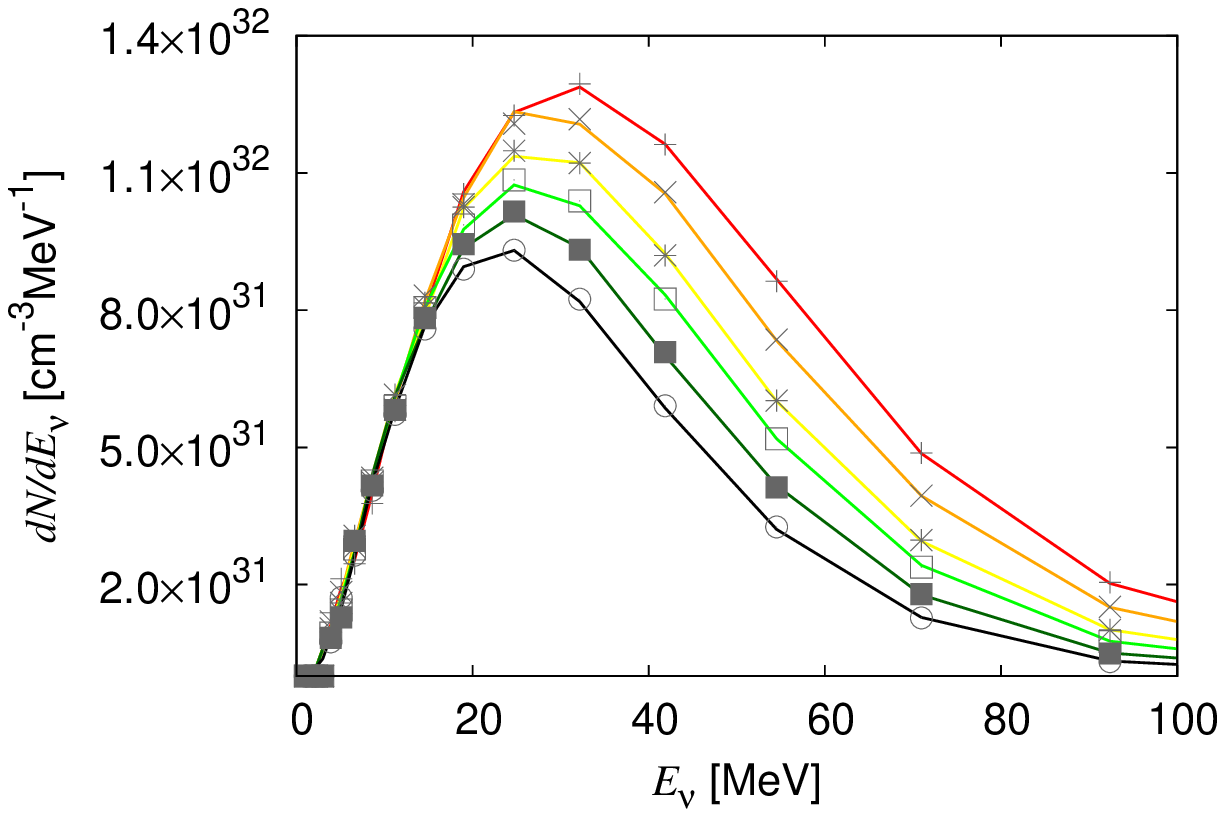}{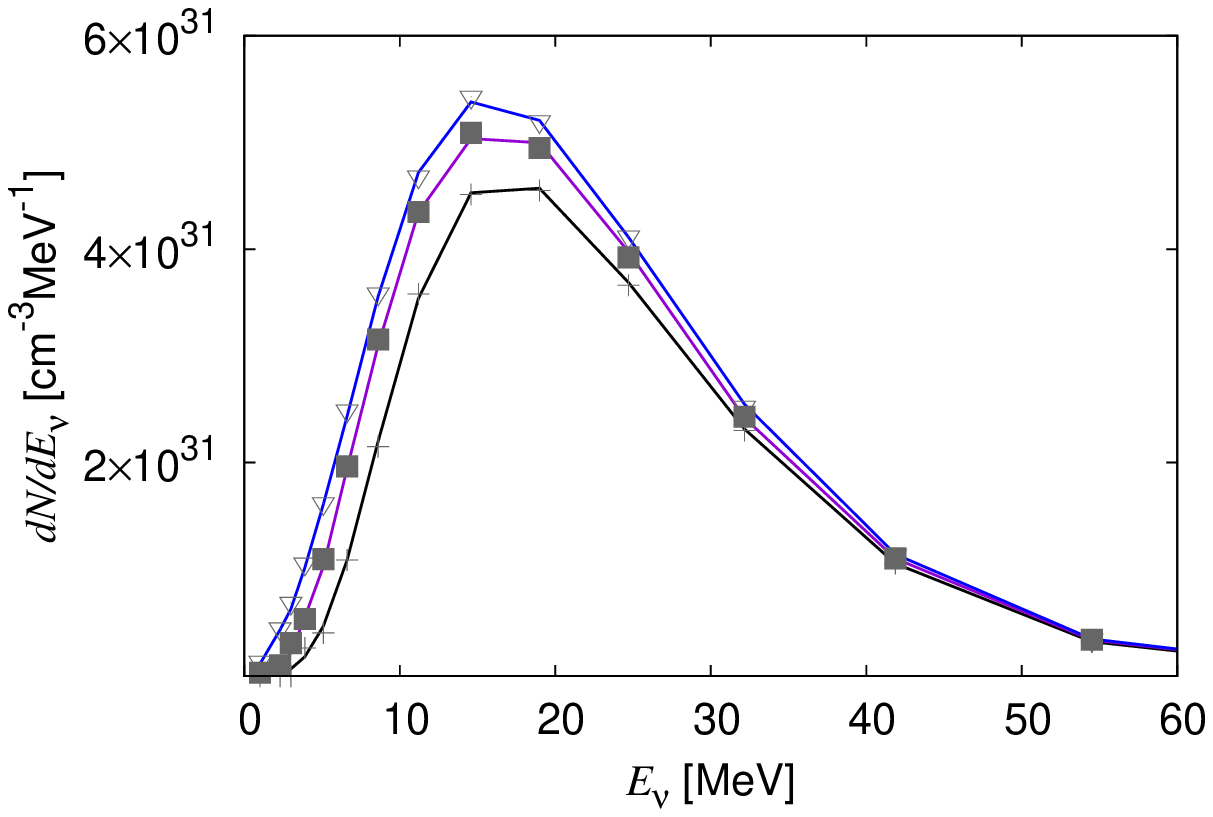}
  \plottwo{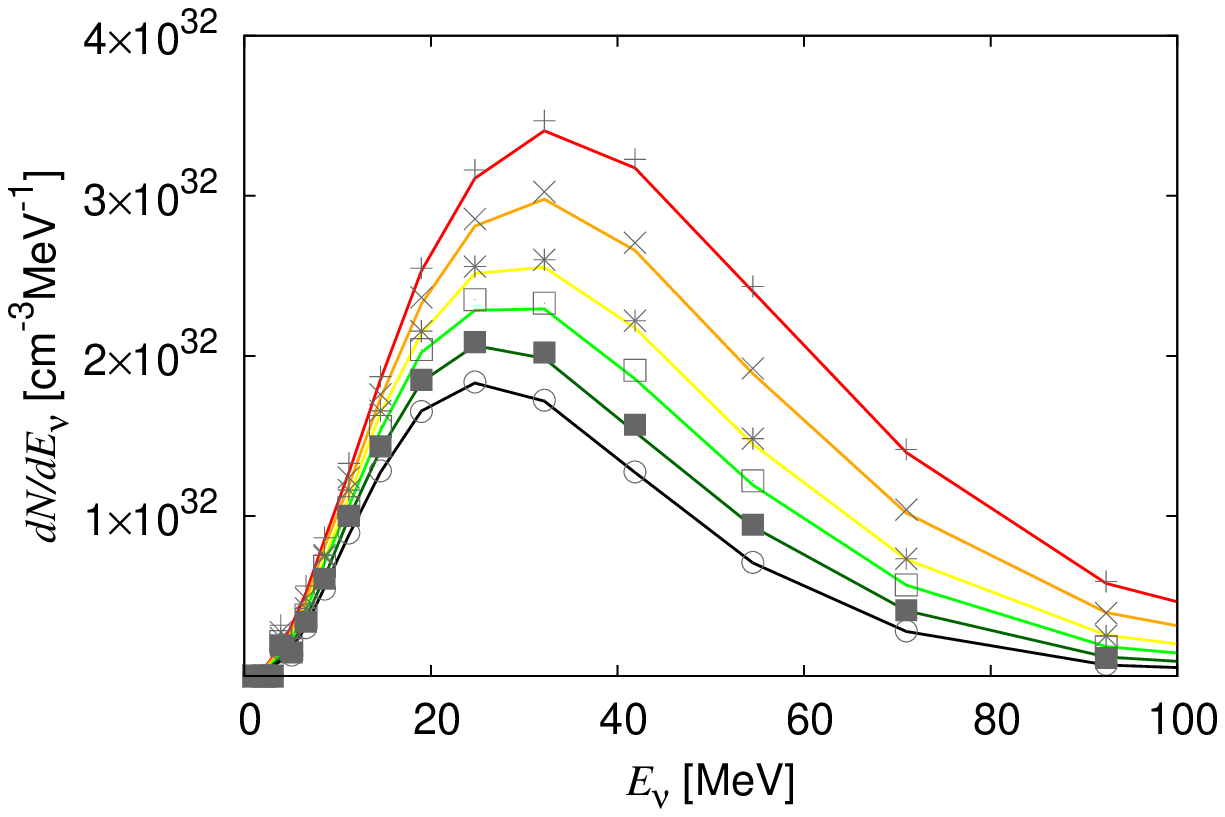}{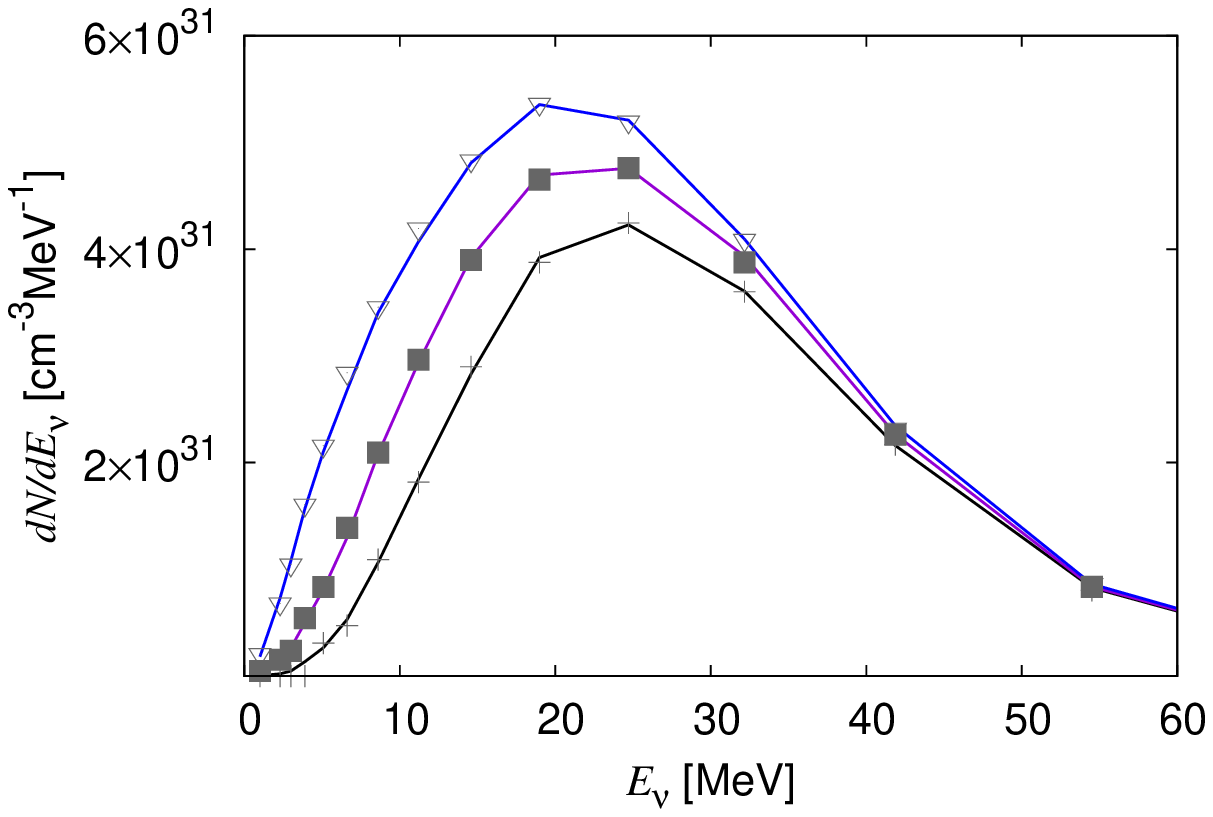}
\caption{The comparison of the energy spectra of $\nu_e$'s (top), $\bar{\nu}_e$'s (middle) and $\nu_x$'s (bottom) between the MC code and the finite-difference Boltzmann solver by \cite{Nagakura:2014nta} for some selected radii in region I (left) and II (right). Color lines represent the MC results and gray symbols show the results by the Boltzmann solver. In the left panels, different lines correspond to different radii and the scattering angle is fixed to $\cos{\theta_\nu}=0.973$, whereas in the right panels, the radius is fixed to $r$ = 34 km and the scattering angle is varied. \label{compair}}
\end{figure*}

\begin{figure}[htbp]
\center
\epsscale{1.4}
\plotone{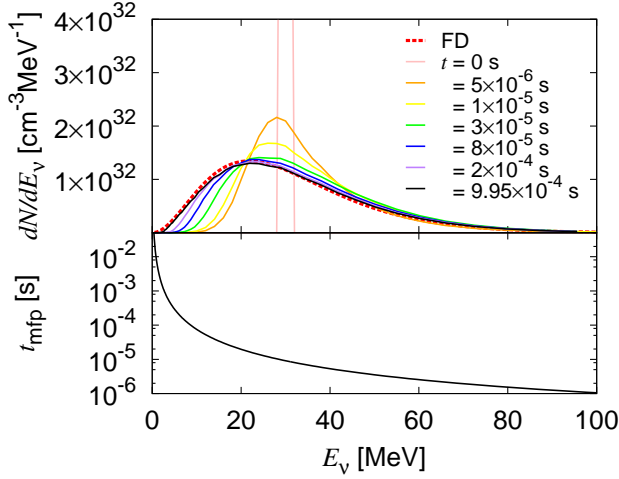}
\caption{The thermalization of neutrino spectrum by neutron recoils. In the upper half, the solid lines present the spectra at different times and the red dotted line gives the Fermi-Dirac distribution $f_{\rm{eq}}$ with $T = 9.96$ MeV and $\mu_\nu = -1.75$ MeV expected after thermalization. The lower half exhibits the mean free time of neutrinos as a function of the neutrino energy. \label{therm}}
\end{figure}

\begin{figure}[htbp]
 \center
 \epsscale{1.2}
  \plotone{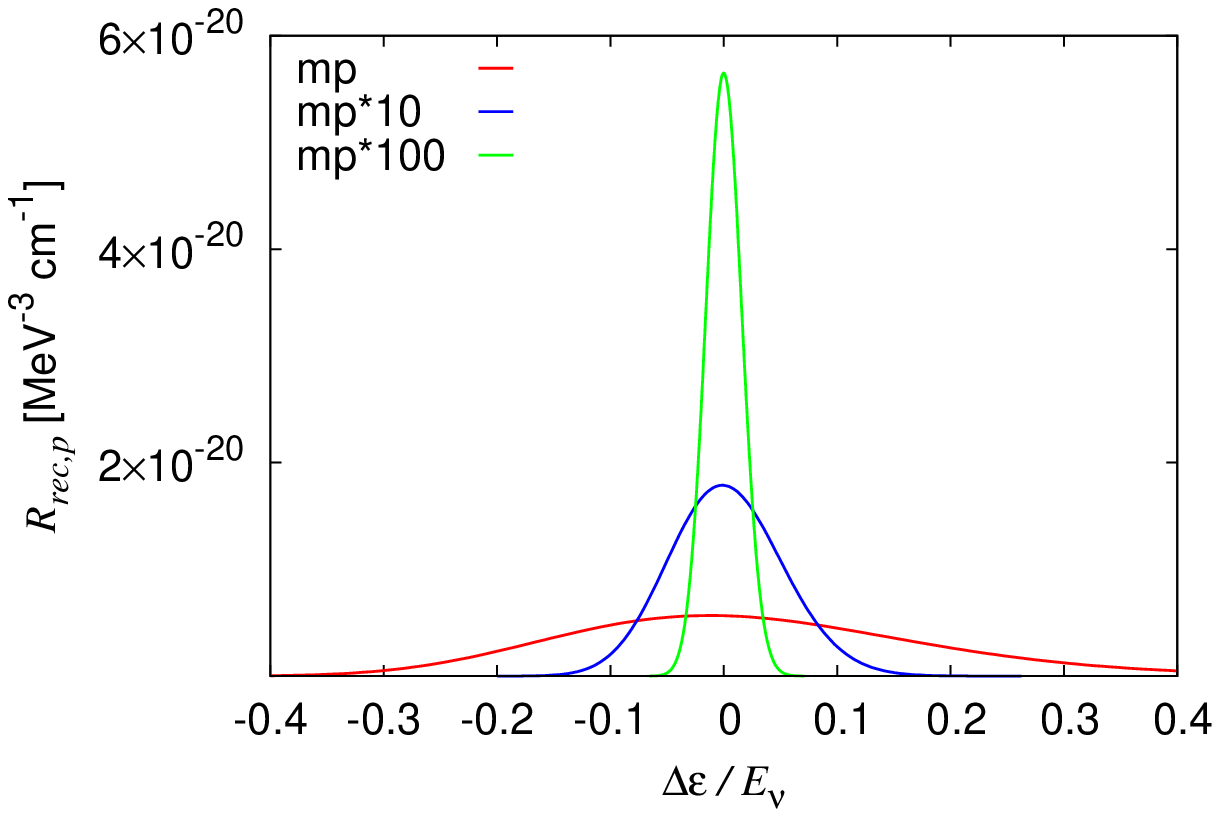}
  \plotone{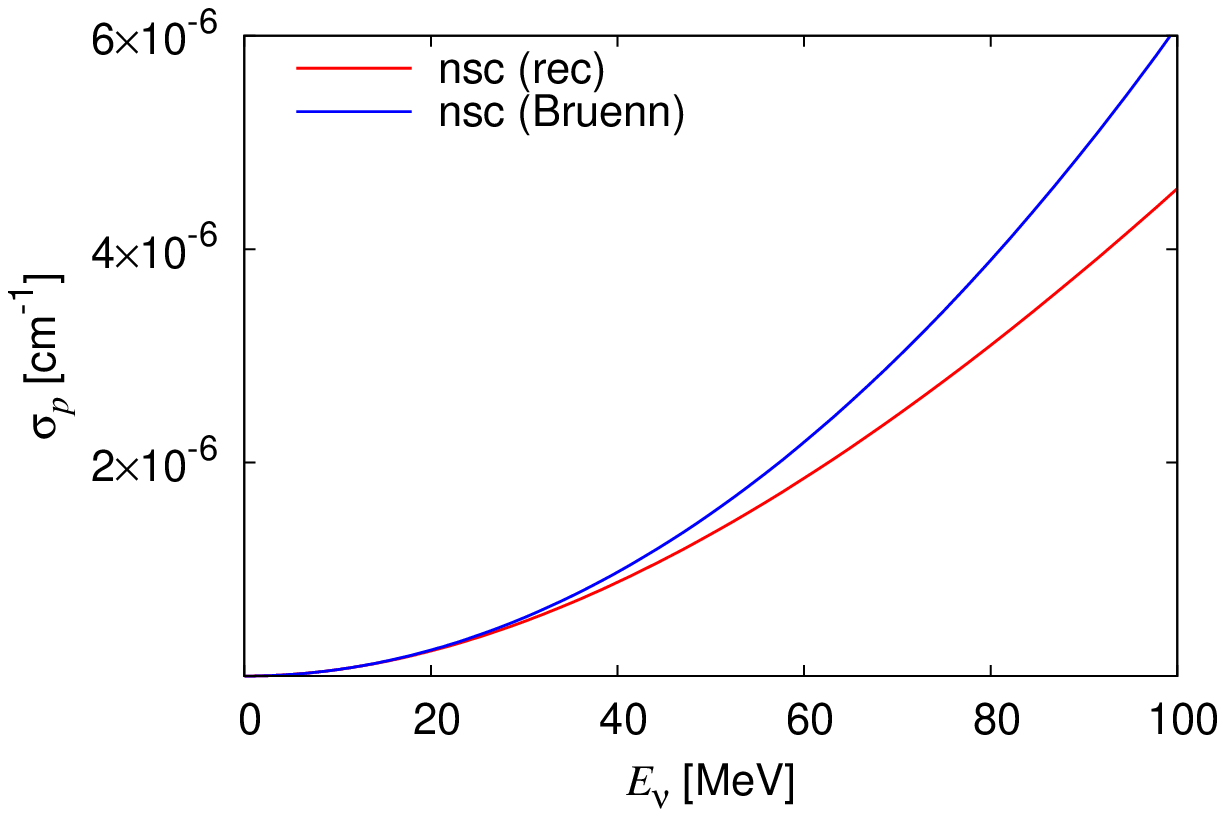}
  \plotone{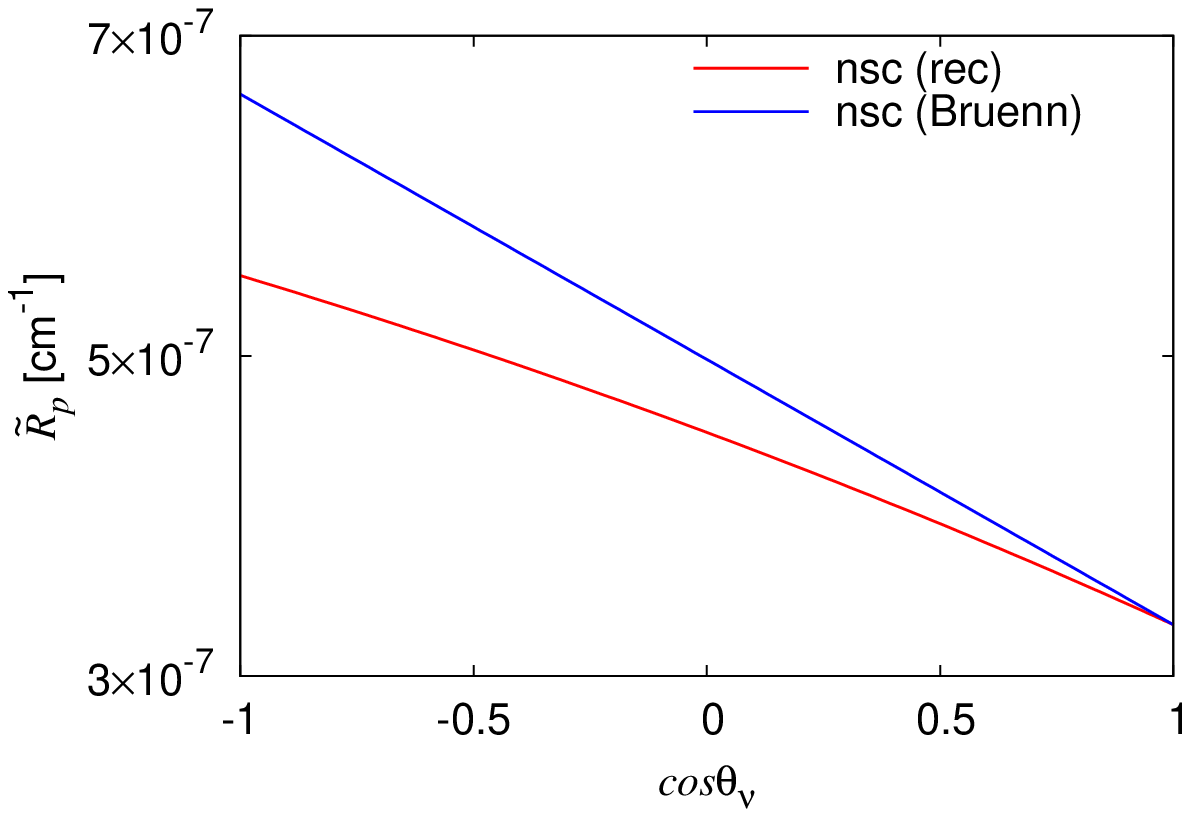}
\caption{Top: the proton scattering rate as a function of proton mass: $m_p$ (red), $10\times m_p$ (blue) and $100\times m_p$ (green). The horizontal axis is the ratio of the lost energy to the initial energy. Middle: the cross sections of the proton scattering with (red) and without (blue) recoils as a function of neutrino energy.
Bottom: the angle dependence of the proton-scattering rates at $E_\nu = 40$ MeV with (red) and without (blue) recoils. \label{ef_rec}}
\end{figure}

\section{Code validation} \label{ch3}

In this section we present some of the test calculations we conducted for the validation of our MC code.
We first compare the results obtained with our MC code and those with another Boltzmann solver based on discretization \citep{Nagakura:2014nta,Nagakura:2016jcl,2019ApJ...878..160N} in Section~\ref{sub:compair}.
The numerical treatment of the detailed balance in the neutrino-nucleon scattering, a key ingredient in this paper, is then validated in the computation of the thermalization of neutrino spectrum via this process in a single spatial zone in Section~\ref{sub:thermalization}.

%We develop a new code for the neutrino transport using the MC method.
%For the validation of our code, we perform several test calculations.
%Two of them are described in this section.

\subsection{Comparison with the finite-difference Boltzmann solver} \label{sub:compair}

We validate our MC code with another Boltzmann solver developed by \cite{Nagakura:2014nta,Nagakura:2016jcl,2019ApJ...878..160N}, which is a finite-difference code based on the $S_N$ method.
We take a similar strategy to that in \cite{2017ApJ...847..133R}: we employ a snapshot at 100 ms after bounce taken from our realistic one-dimensional dynamical SN simulation with $M_{\text{ZAMS}}$ = 11.2 $M_\odot$ \citep{2019ApJS..240...38N}; fixing the matter distribution so obtained, we run the two neutrino transport codes to obtain a steady neutrino distribution.
Note that the same background model is used for the later studies.
Top three panels in Figure~\ref{hydro} show the radial profiles of density, temperature and electron fraction in this model.
We focus on two regions: region I ($r$ = 20 -- 25 km) and region II ($r$ = 28 -- 34 km) painted in yellow.
In the former region, neutrinos are nearly in thermal equilibrium with matter, whereas in the latter region they get gradually out of equilibrium as the density decreases and their distribution starts to become anisotropic.

The set of neutrino reactions employed in this comparison is referred to ``base'' in Table~\ref{reac_MC}.
Note that the nucleon recoil is not included.
We deploy $2\times10^6$ sample particles and adopt the time step of  $dt_{\rm{f}}= 10^{-7}$ s, which is the same as the time step for updating the neutrino distribution function in this case (see Appendix~\ref{appendix2}).
We adopt exactly the same spatial grid as employed in the SN simulation and assume that hydrodynamical quantities are constant in each cell.
In order to set the inner and outer boundary conditions, we introduce ghost cells both inside and outside the active region and deploy sample particles uniformly there according to the distribution functions imposed at the boundaries.
Turning off all the interactions with matter, we follow the motions of these sample particles in the ghost cells to make the fluxes at the boundaries as close to the prescribed values as possible.
%It is clear that neutrinos, which come from the active region, never go back again.
%Moreover, we take the width of the boundary regions longer than the distribution length $l_{\rm{f}}$ and reset conditions of the boundary regions at the beginning of each time step.}

%Although it is not trivial to decide the system settles down to a stead-state since neutrino distributions fluctuate with time due to statistical errors in the MC simulation, we define it as the state, 

We follow the time evolution of neutrino radiation field by MC simulations until the system settles down to a nearly steady state, in which the total number of sample particles do not change more than 0.5 \% from a certain value for the total number of sample particles.
We then take the average over 8,000 time steps ($8\times10^{-4}$ s) after the steady-state is achieved to reduce the statistical error, and evaluate the number spectra of neutrinos from the mean distribution function.
Note that neutrinos with $E_\nu \gtrsim 5$ MeV experience scatterings with nucleons more than 10 times during this period.
This may be understood from the total mean free path for the nucleon-scattering\footnote{Note that we use the exact reaction rate $R_{\rm{rec}}$ for the cross sections of nucleon scattering $\sigma_N$ in the bottom panel of Figure~\ref{hydro}.} in the bottom panel of Figure~\ref{hydro}.

Figure~\ref{compair} shows the comparison of the energy spectra $dN(r,E_\nu)/dE_\nu$:
\begin{eqnarray}
 \frac{dN\left(r,E_\nu \right)}{dE_\nu} = \frac{1}{\left(2\pi \hbar c\right)^3} \int 2\pi E^2_\nu f\left(r,E_\nu,\theta_\nu\right) d\cos{\theta_\nu},\ \ \ 
\end{eqnarray}
for $\nu_e$'s (top), $\bar{\nu}_e$'s (middle) and $\nu_x$'s (bottom) between the MC code and the finite-difference Boltzmann solver.
The left panels show the results in the region I.
Color lines correspond to the results of the MC calculation for $\cos{\theta_\nu} = 0.973$ at different radii.
We use the same energy and angle grids as those employed by the finite-difference Boltzmann solver to facilitate comparisons.
Gray symbols present the results obtained with the finite-difference Boltzmann solver.
We find a good agreement between the two methods.

In the right panels, on the other hand, we pick up the neutrino spectra at $r$ = 34 km in the region II.
Different colors denote the different cosines of angles $\cos{\theta_\nu}$.
One can see that the angular distributions of neutrinos start to become forward-peaked with $\nu_x$ being the most anisotropic as expected.
The neutrino spectra given by our MC code are again in an excellent agreement with those by the finite-difference Boltzmann solver in this bit outer region.

\subsection{Thermalization by nucleon recoils} \label{sub:thermalization}
In this paper, we focus on the effects of nucleon recoils, particularly the thermalization of neutrinos.
In so doing, the detailed balance should be satisfied in the numerical simulations:
\begin{eqnarray}
  &&R_{\rm{rec}}\left(E_\nu,E^\prime_\nu,\cos{\theta_\nu} \right)f_{\rm{eq}}\left(E_\nu\right)\left(1-f_{\rm{eq}}\left(E^\prime_\nu\right)\right) \nonumber \\
  &&\ \ \ = R_{\rm{rec}}\left(E^\prime_\nu,E_\nu,\cos{\theta_\nu} \right)\left(1-f_{\rm{eq}}\left(E_\nu\right)\right)f_{\rm{eq}}\left(E^\prime_\nu\right). \ \ 
\end{eqnarray}
%In this expression, $R^\prime_{\rm{rec}}$ is the integrated reaction rate over the angle $\psi$:
%\begin{eqnarray}
%  &&R^\prime_{\rm{rec}}\left(E_\nu,E^\prime_\nu\right) \nonumber \\
%  &&\ \ \ \equiv \int^1_{-1} 2\pi E^{\prime2}_\nu R^\prime_{\rm{rec}} \left(E_\nu,E^\prime_\nu,\cos{\psi}\right) d\cos{\psi}.
%\end{eqnarray}
This is ensured simply by calculating the reaction rates for $E_\nu \le E^\prime_\nu$ and obtaining those for the other case $E^\prime_\nu > E_\nu$ from the former so that the detailed balance is guaranteed.
We tabulate the reaction rates obtained for the thermodynamical conditions encountered in the matter background.
The detailed procedure is described in Appendix \ref{appendix}.

Ignoring the spatial dependence, we perform a one-zone calculation with $T=9.96$ MeV and the chemical potential of neutrons, $\mu_n = 921$ MeV, this time.
We follow the thermalization of neutrino spectra only by neutrino-neutron scatterings in this test.
We inject sample particles with the monochromatic energy, $E_\nu$ = 30 MeV, as an initial condition.
%We assume that all sample particles have 30 MeV, initially.
Figure~\ref{therm} shows the time evolution of the neutrino spectrum.
Different colors correspond to different time steps.
The expected thermal spectrum (red dotted) is obtained from the Fermi-Dirac distribution $f_{\rm{eq}}$ as 
\begin{eqnarray}
\frac{dN\left(E_\nu\right)}{dE_\nu} = \frac{1}{\left(2\pi \hbar c\right)^3}\frac{4\pi E^2_\nu}{1+\exp{\left(\frac{E_\nu-\mu_\nu}{T}\right)}}.
\end{eqnarray}
Since the total number of neutrinos $N$ is conserved in this calculation, the chemical potential of neutrinos $\mu_\nu$ is determined by $N$ and $T$.
In this test, we set $N=10^{28}$, which leads to $\mu_\nu= -1.75$ MeV.
We find that the neutrino spectrum approaches this distribution indeed and they are in good agreement with each other at the end ($t = 9.95 \times 10^{-4}$ s) (see the red dotted and black solid lines in Figure~\ref{therm}).
This lends confidence to our treatment of the nucleon scattering for the detailed balance.
%This implies that the detailed balance is satisfied in our treatment of nucleon scatterings.
We also give in the bottom panel of the same figure the mean free time of neutrinos $t_{\rm{mfp}}$ for the neutrino-neutron scattering:
\begin{eqnarray}
t_{\rm{mfp}} \equiv \frac{\lambda_{n}}{c} = \frac{1}{\sigma_n c}.
\end{eqnarray}
The exact reaction rate $R_{\rm{rec},n}$ is used for the cross section $\sigma_n$ in the evaluation.
We find that the computation time is long enough to guarantee the thermalization except at the lowest end of energies, where the scattering occurs only rarely.

%%%%%%%%%%%%%%%%%%%%%%%%%%%%%%%%

\begin{figure}[htbp]
\center
\epsscale{1.2}
\plotone{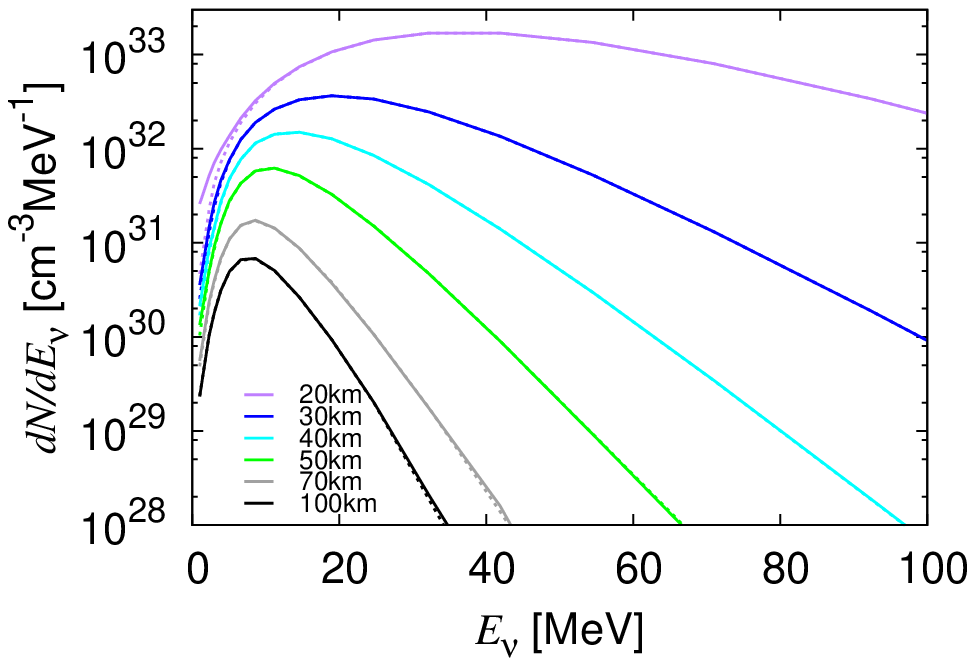}
\plotone{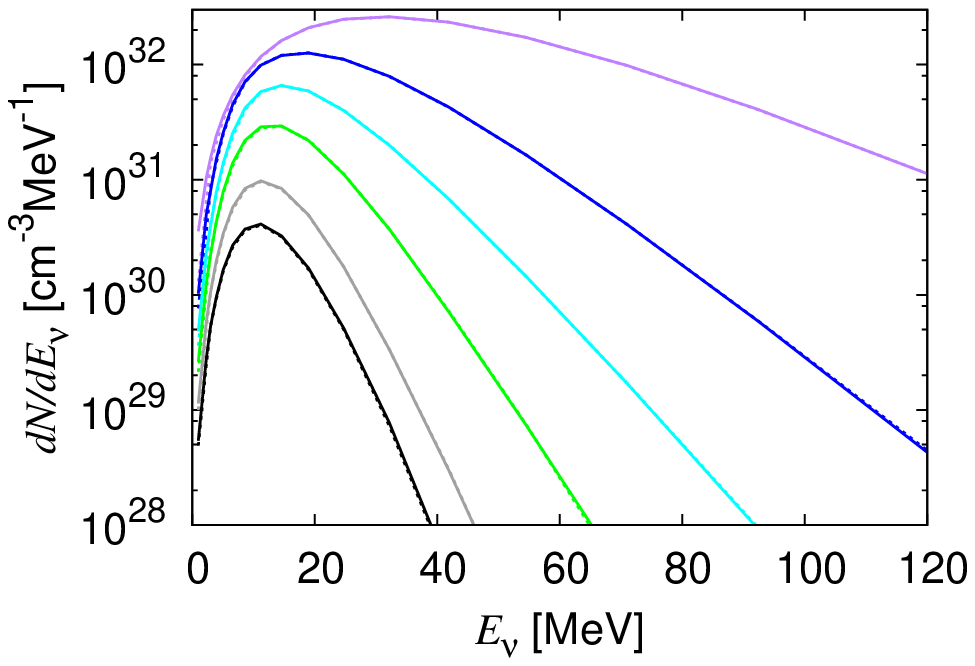}
\plotone{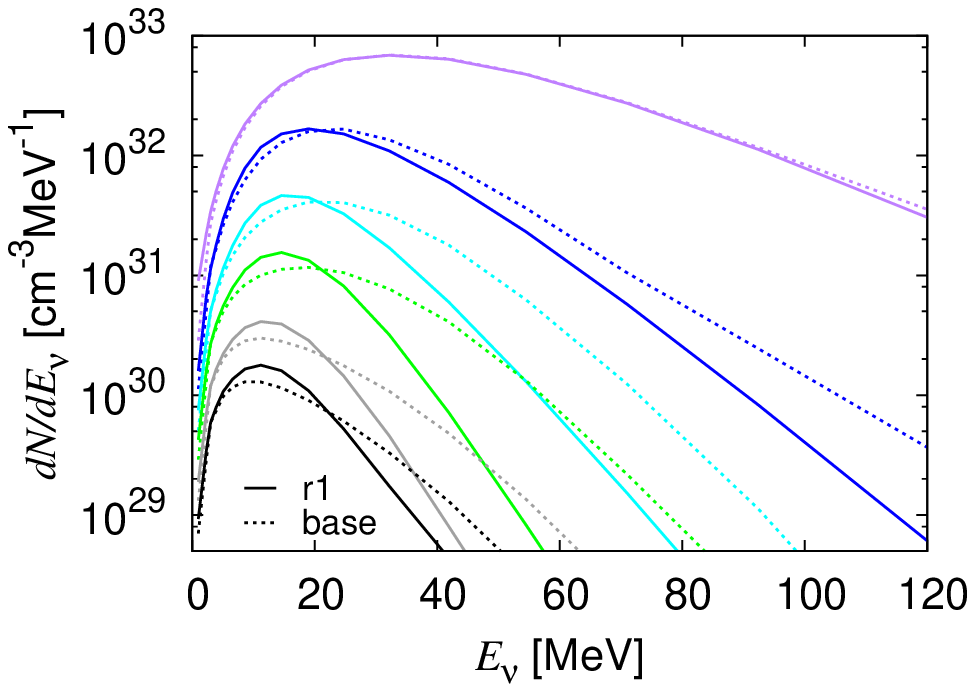}
\caption{The energy spectra of neutrino for the ``base'' (dotted) and ``r1'' (solid) sets of neutrino reactions (see Table~\ref{reac_MC}). Line colors denote the radii. The top, middle and bottom panels show the spectra of $\nu_e$'s, $\bar{\nu}_e$'s and $\nu_x$'s, respectively.\label{spe:rec}}
\end{figure}

%\section{Nucleon recoils and their implementation method} \label{ch4}
\section{Impacts of nucleon recoils on neutrino spectra} \label{ch4}

We apply the MC code to the thermalization of energy spectra as neutrinos propagate outwards in the post-shock region.
We pay particular attention to the relative importance of various processes including the nucleon recoil for different neutrino flavors.
%investigate the effects of nucleon recoils by the steady-state calculations using the developed MC code in Section~\ref{subch:rec} and discuss about which reaction is dominant in the thermalization of neutrino spectra.
%In Section~\ref{subch:namashi}, \textcolor{black}{we adopt same energy grids as Boltzmann solver to our MC code and make the distribution uniform in each energy bin at the end of each time step, which is same situation as the discretized method. 
%By the steady-state calculations, we estimate the deviation from the correct neutrino spectra.}
%We also suggest a new way to reconstruct neutrino spectra in the discretized method.

%\subsection{The effect of nucleon recoils} \label{subch:rec}

\subsection{Iso-energy limit of nucleon scattering}
Before looking into the individual contributions of different processes to the thermalization of neutrino spectra in detail, it may be worth to see the iso-energy limit of the nucleon scattering, which was derived by \cite{1985ApJS...58..771B} and was employed in most of SN simulations in the past. %, because the nucleon mass is much larger than the average neutrino energies and, as a result, the energy exchange per scattering between neutrino and nucleon is very small indeed.
The Bruenn rate (eq.~(\ref{bruenrate})) can be derived from the generic expression for the non-isoenergetic scattering (eqs.~(\ref{R_rec})-(\ref{R_rec_f})) by taking a limit of $m_N \rightarrow \infty$ and $\Delta \epsilon/E_\nu \rightarrow 0$.

The top panel of Figure~\ref{ef_rec} shows the dependence on the proton mass of the reaction rate for the proton scattering.
The vertical axis is the reaction rate $R_{\rm{rec}, p}$ and the horizontal axis is the ratio of the energy change to the initial energy, $\Delta \epsilon/E_\nu$.
It is clear that as the proton mass increases, the energy exchange becomes smaller, making the reaction rate more sharply peaked at $\Delta \epsilon/E_\nu = 0$, the iso-energetic scattering limit.
Note that in these calculations of $R_{\rm{rec}, p}$ we modify the chemical potential of protons so that the number density should be unchanged.

In addition to the energy re-distribution, the effect of proton recoils is the reduction of the reaction rate at high energies and/or at backward scattering-angles as shown in the middle and bottom panels of Figure~\ref{ef_rec}, respectively, for $T = 5.85$ MeV, $\rho = 10^{12}\ \rm{g/cm^3}$ and $\mu_p = 907$ MeV.
We find that the latter effectively modifies the angular dependence of the nucleon scattering, making it less backward-peaked.

%they are reduced at the high energy, especially.
%The reaction rate for protons $\tilde{R}_{\rm{rec}}$ are calculated at the same condition under the assumption that the initial energy $E_\nu = 40$ MeV and we find that they are reduced in the case of backward scatterings. 

\subsection{Sensitivity of neutrino spectra on recoils in the nucleon scattering} \label{subch:thermal_neutrino}
%We now move on to the investigation of the thermalization of neutrino spectra.
We assess the impact of nucleon recoils by comparing the energy spectra in MC simulations with/without the recoils on a realistic CCSN matter background.
We run the MC code to obtain steady-state solutions of the neutrino transport on the static matter background given by the same progenitor model employed in the code validation (see Figure~\ref{hydro}).
The inner and outer boundaries are put at 20 and 100 km, respectively.
%The inner boundary is necessary as the optically thick region is hard to deal with by the MC method.
%It is actually unnecessary, since neutrinos are in equilibrium with matter anyway.
%Outside the outer boundary, on the other hand, neutrinos move freely and their spectra no longer change essentially.
The neutrino fluxes coming in from these boundaries are obtained automatically by setting the neutrino distribution functions on the ghost mesh points to the ones derived from the SN simulation.

As the first comparison, we adopt two sets of neutrino reactions: ``base'' and ``r1'' given in Table~\ref{reac_MC}.
In the r1 set, the nucleon recoil is taken into account in addition to the base set.
For both cases of calculations, we use $2\times10^6$ sample particles and take $dt_{\rm{f}} = 10^{-7}$ s for the distribution time. 

\begin{figure}[htbp]
\centering
\epsscale{1.2}
\plotone{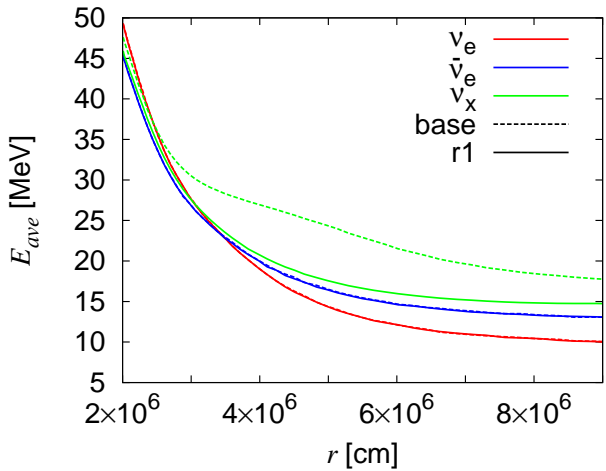}
\caption{The radial profiles of the average energies of $\nu_e$'s (red), $\bar{\nu}_e$'s (blue) and $\nu_x$'s (green). Solid and dotted lines correspond to the ``r1'' and ``base'' sets of neutrino reactions, respectively (see Table~\ref{reac_MC}). \label{ave_ene}}
\end{figure}

Figure~\ref{spe:rec} shows the energy spectra of neutrino number densities obtained in the two calculations.
Colors denote the radii, at which the spectra are evaluated, and solid and dotted lines show the results for the r1 and base sets, respectively.
%The number spectrum is defined as
%\begin{eqnarray}
%N\left(r,E_\nu \right) = \frac{1}{\left(2\pi \hbar c\right)^3} \int 2\pi E^2_\nu f\left(r,E_\nu,\theta_\nu\right) d\cos{\theta_\nu}.
%\end{eqnarray}
The spectra of $\nu_e$'s (top) and $\bar{\nu}_e$'s (middle) do not change by the inclusion of the nucleon recoil, whereas high-energy $\nu_x$'s are depleted and low-energy ones are increased due to down-scatterings by nucleons (bottom).
As a result, the average energy of $\nu_x$'s is reduced by $\sim$~15\% at the outer boundary as shown in Figure~\ref{ave_ene}.
Note that the maximum difference is $\sim$~30\% at $r\sim40$ km.
The number density of $\nu_x$'s is also decreased by $\sim$~7\% at the outer boundary.
This is due to the opacity reduction caused by the nucleon recoil itself as well as by the decrease of average energy.

\begin{figure*}[htbp]
  \epsscale{1.0}
  \plottwo{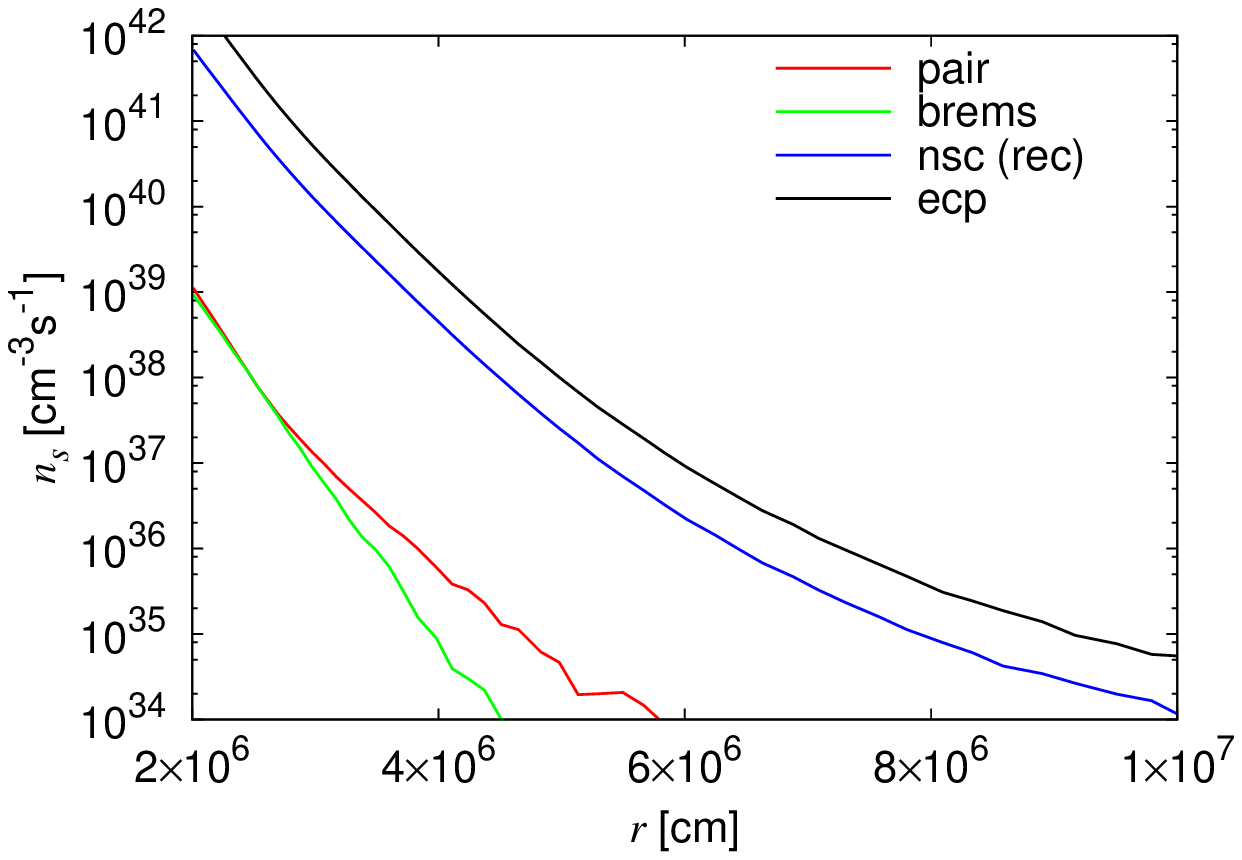}{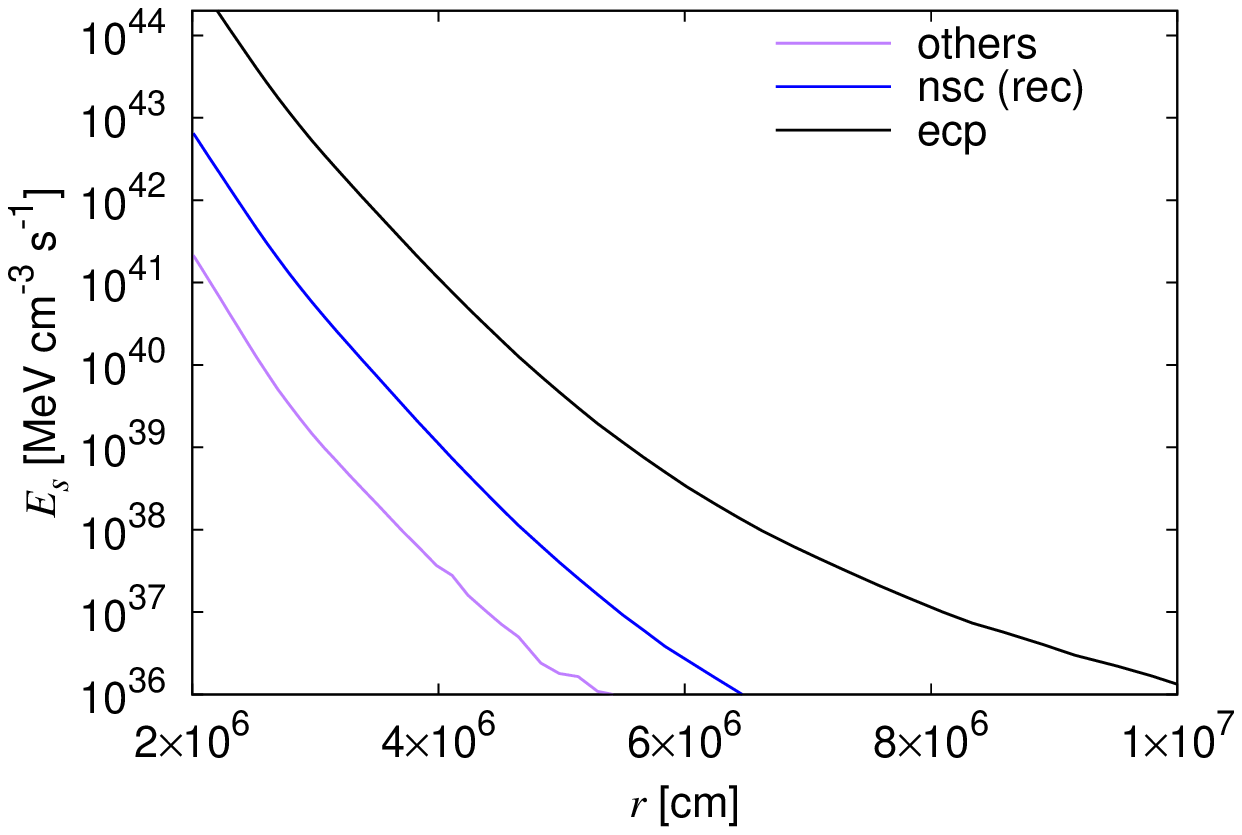}
  \plottwo{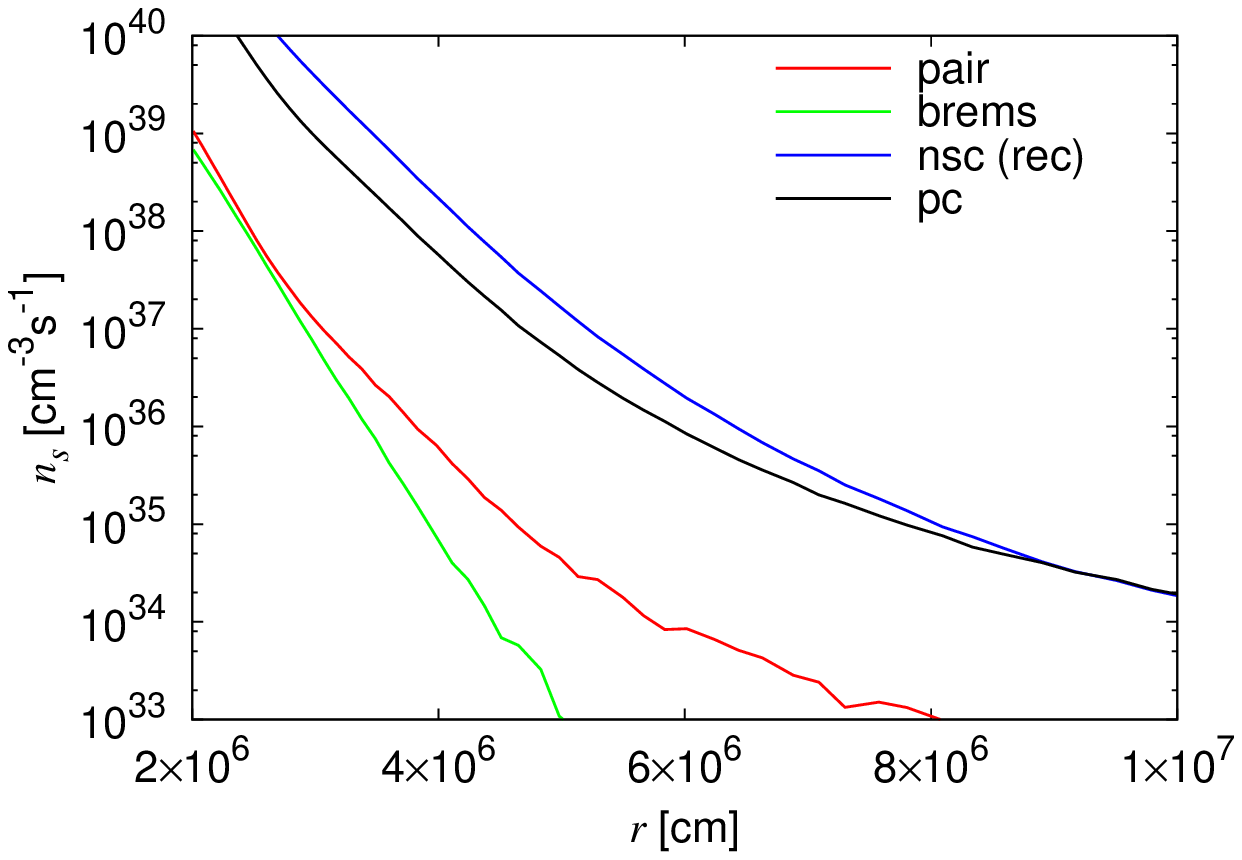}{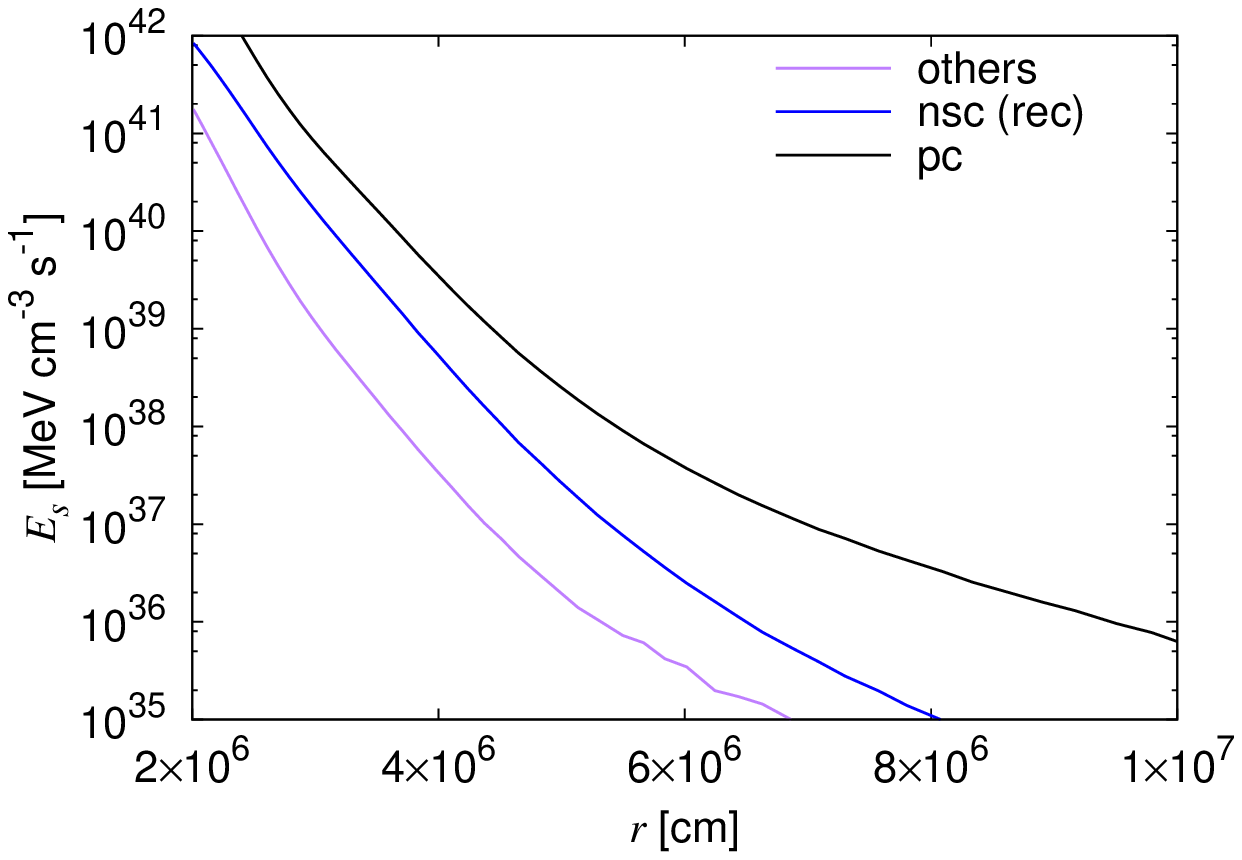}
  \plottwo{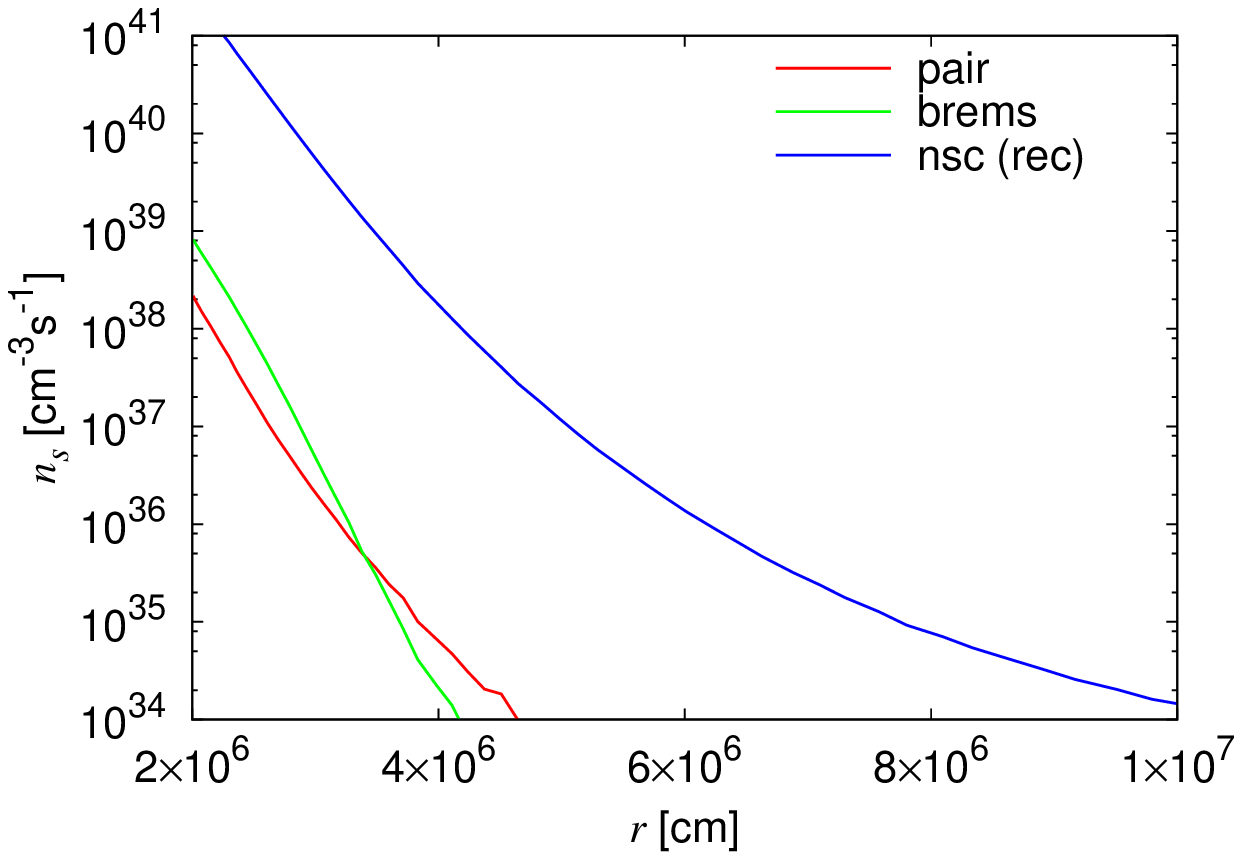}{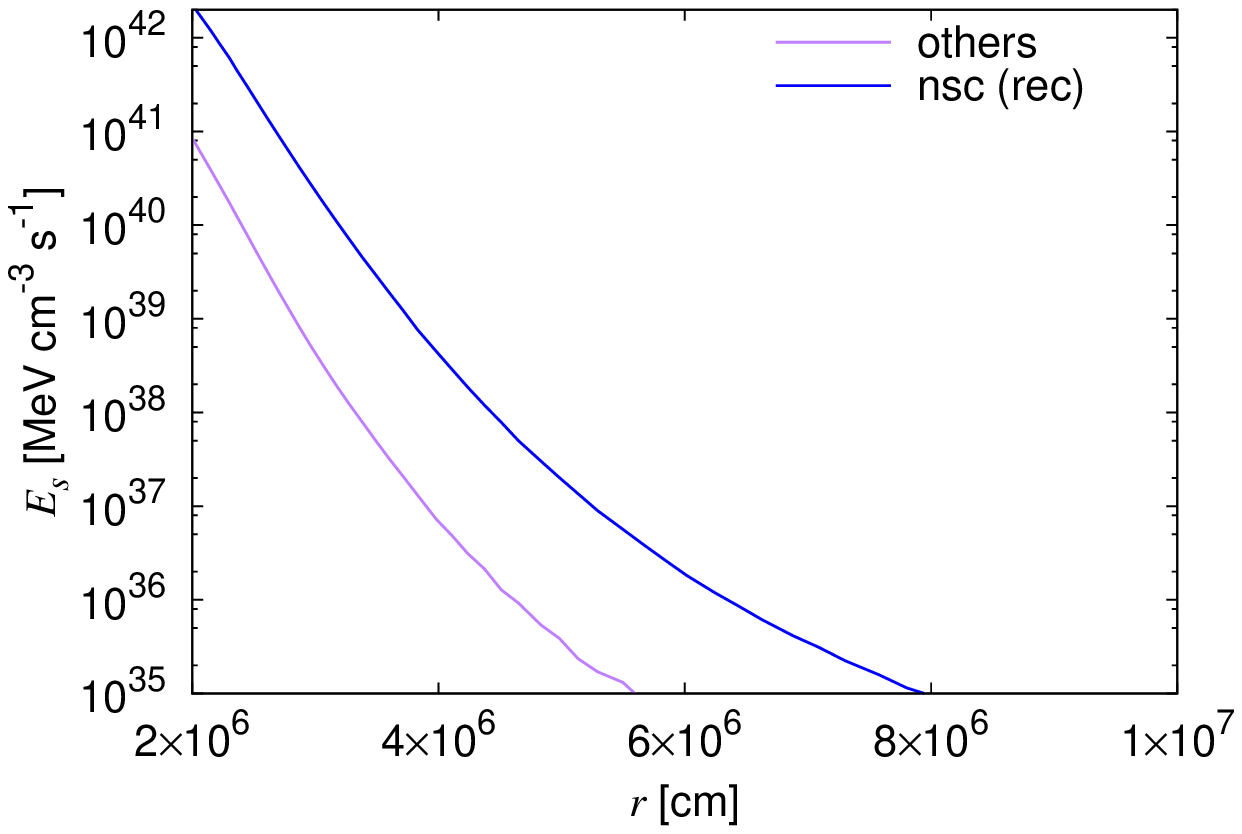}
\caption{Left: the radial profiles of the number of neutrinos, which experience interactions with matter per unit time and volume on each neutrino reaction for $\nu_e$'s (top), $\bar{\nu}_e$'s (middle) and $\nu_x$'s (bottom). Right: the radial profiles of the energy exchanged between neutrino and matter on each neutrino reaction. In the right panels, the pair-annihilation and bremsstrahlung are put together into ``others''. \label{reac:rec}}
\end{figure*}

In order to understand the different responses to the inclusion of the nucleon recoil among different flavors, we show the rates per volume for different reactions as a function of radius in the left panels of Figure~\ref{reac:rec}.
Line colors denote the different reactions.
The vertical line shows the number of neutrinos, which experience each neutrino reaction per unit time and volume, denoted as $n_s$.
One finds in the top panel that the electron capture dominates the other reactions for $\nu_e$'s.
This is the reason why the spectrum is not changed by the inclusion of the nucleon recoil.
Note that the number of nucleon scatterings itself is smaller than that of electron captures by a factor of $\sim$ 5.
The dominant reaction for $\nu_x$'s, on the other hand, is the nucleon scattering in the absence of charged-current reactions (see the bottom panel).
As a result, the spectrum is pinched by the inclusion of the nucleon recoil.
For $\bar{\nu}_e$'s (middle), the number of nucleon scatterings is larger than those of the other reactions.
Although this seems to contradict at first glance with the previous result that the spectrum of $\bar{\nu}_e$'s is not affected by the nucleon recoil, this is simply due to the small energy exchange in the nucleon scattering.

The right panels of Figure~\ref{reac:rec} demonstrates this.
They show the energies exchanged between neutrino and matter for different reactions. 
The vertical axis is the exchanged energy per unit time and volume and denoted by $E_s$.
%The exchange energy by nucleon scatterings is defined as $|E^\prime_\nu-E_\nu|$ and those by absorption and emission are defined as the neutrino energy of absorbed or emitted sample particles.
In the figure, the pair-annihilation and bremsstrahlung are put together into ``others''.
We find for $\nu_e$'s (top) and $\nu_x$'s (bottom) that the orders of lines in the right panels are unchanged from those in the corresponding left panel.
For $\bar{\nu}_e$'s (middle), on the other hand, the positron capture is dominant over the nucleon scattering in terms of the energy exchange although the opposite is true for the reaction rates.
This is, as mentioned above, due to the small energy exchange in the individual scattering on nucleons.
As a result, the nucleon recoil affects the spectrum of $\nu_x$'s but not of $\bar{\nu}_e$'s.
Note that our result is qualitatively consistent with the result in \cite{Keil:2002in}.
%This is consistent with \cite{Keil:2002in}.
%The reduction of neutrino opacities enhances the PNS contraction and increase of temperature, which increase the luminosities of $\nu_e$ and $\bar{\nu}_e$ and the heating behind the shock wave.

%\subsection{Electron/positron scattering} \label{subch:esc}

\begin{figure}[htbp]
\centering
\epsscale{1.1}
\plotone{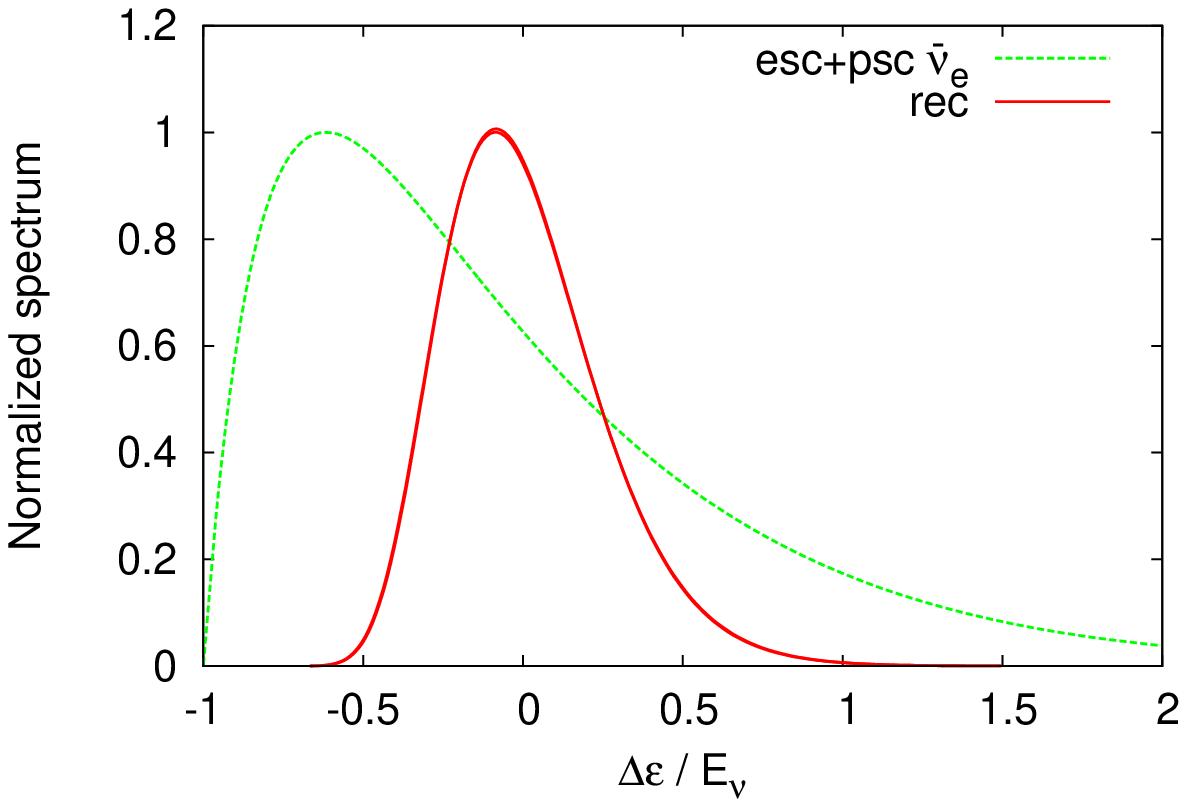}
\plotone{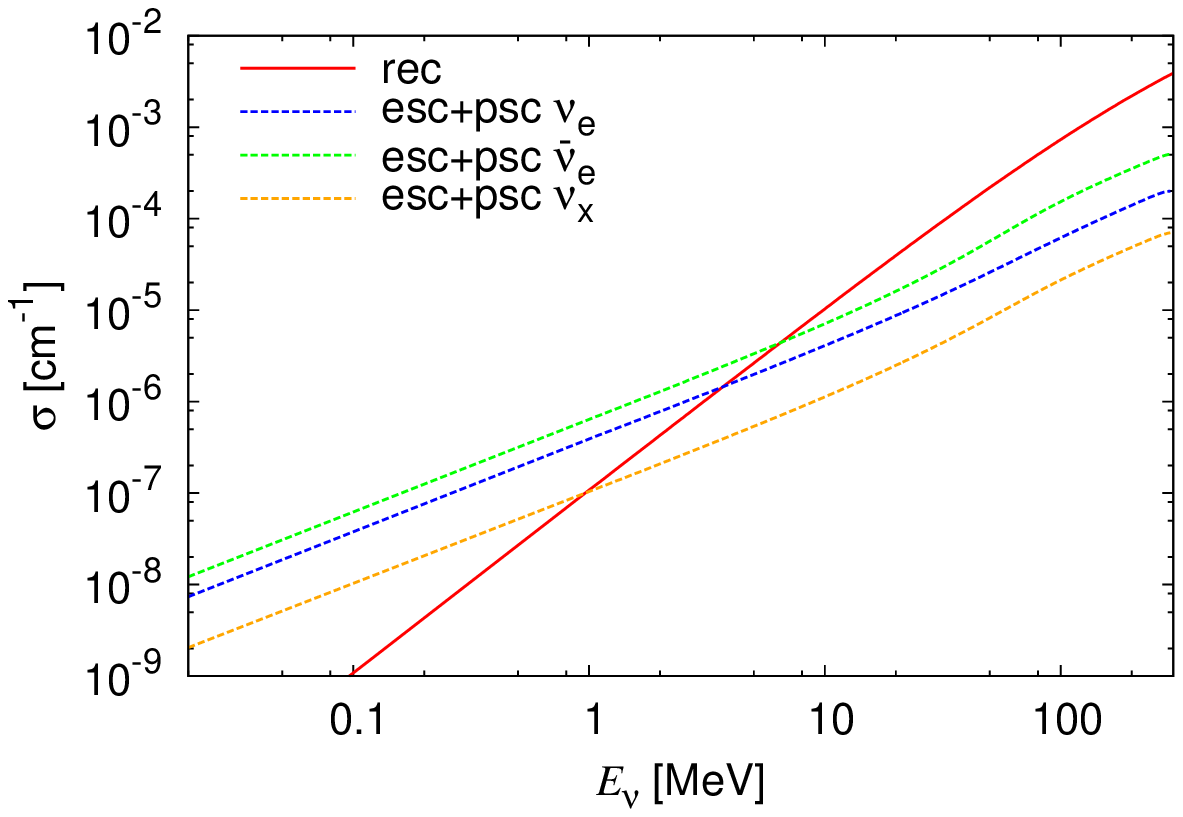}
\caption{Top: the normalized reaction rates of the electron/positron scattering for $\bar{\nu}_e$'s (green) and the nucleon scattering with recoils (red) as a function of the energy change normalized by the initial neutrino energy. Bottom: the total cross sections of the nucleon scattering (red solid) and the electron/positron scattering (dotted) for each neutrino species. \label{rec:esc}}
\end{figure}

We have so far omitted electron/positron scatterings on purpose.
\textcolor{black}{The energy exchange per scattering for electron and positron is much larger than that for nucleon because of the smaller mass of the former, $m_e = 0.511$ MeV.}
In the top panel of Figure~\ref{rec:esc}, we compare the energy exchanges between the two scatterings for the incident-neutrino energy $E_\nu = 25$ MeV and the scattering angle $\cos{\theta}_\nu = -1.0$.
Note that we show the case of $\bar{\nu}_e$'s for the electron/positron scattering.
The vertical and horizon axes are the normalized reaction rate and the ratio of the energy exchange to the incident energy, respectively.
It is clear that the peak of the reaction rate for the electron/positron scattering is dislocated from the iso-energy condition $\Delta \epsilon/E_\nu = 0$ by a large amount, which means that neutrinos give larger energy to electrons/positrons than to nucleons on average.
In the bottom panel of Figure~\ref{rec:esc}, we show the total cross sections for the two scatterings as a function of the incident-neutrino energy.
For the electron/positron scattering, we give them separately for the three neutrino flavors.
We calculate these cross sections at \textcolor{black}{$T=5.85$ MeV, $\rho=1.01\times10^{12}\ \rm{g/cm^3}$, $Y_e = 0.10$, $\mu_p=907$ MeV, $\mu_n=924$ MeV and $\mu_e=19.6$ MeV.}
One finds that the nucleon scattering has larger cross sections at $E_\nu \gtrsim$ a few MeV because of the different energy dependences of the total cross sections: $\sigma \propto E_\nu^2$ for the nucleon scattering whereas $\sigma \propto E_\nu$ for the electron/positron scattering.

We now rerun the MC code, this time with the e1 set of the neutrino reactions given in Table~\ref{reac_MC}, in which the electron/positron scattering is taken into account in addition to the r1 set.
The number of sample particles and the distribution time $dt_f$ are the same as those in the previous calculations.
This run is meant to see the relative importance of the two scatterings in thermalizing the neutrino spectra.

\begin{figure}[htbp]
\epsscale{1.2}
\plotone{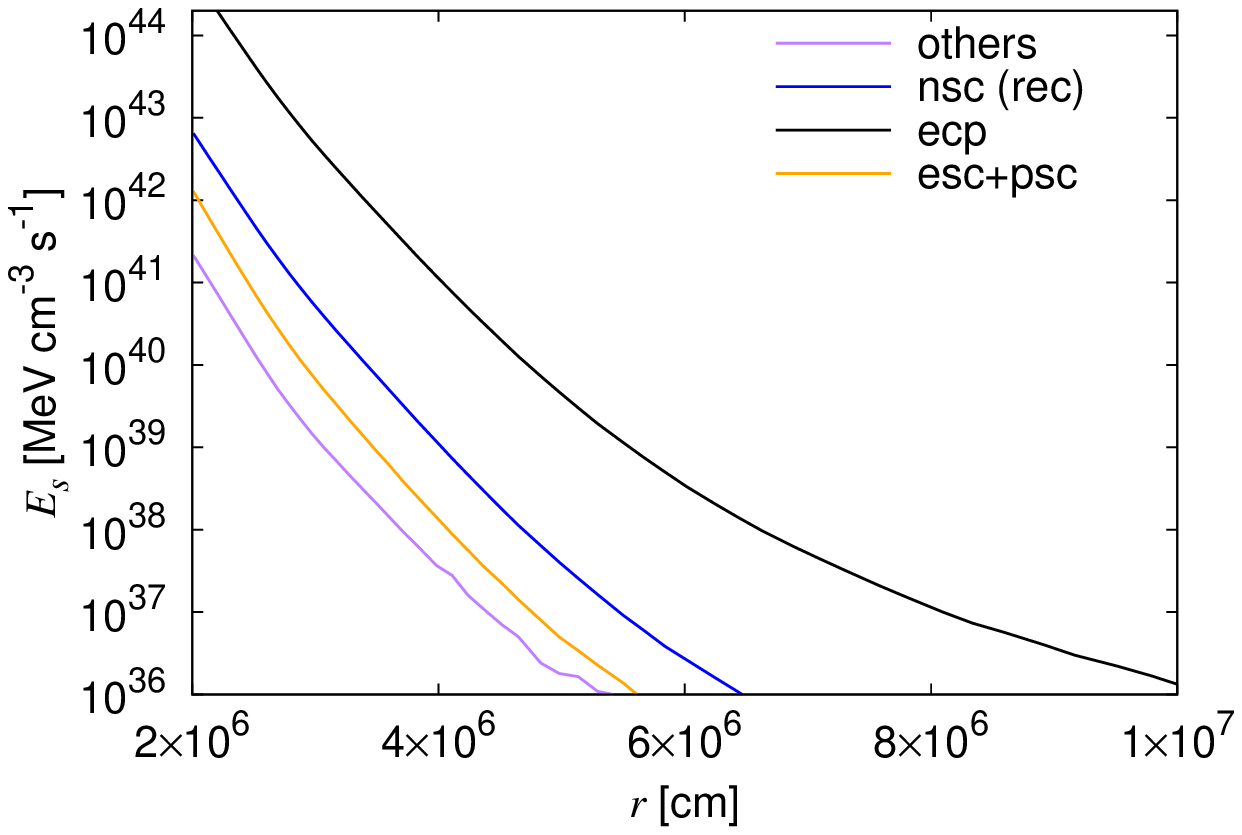}
\plotone{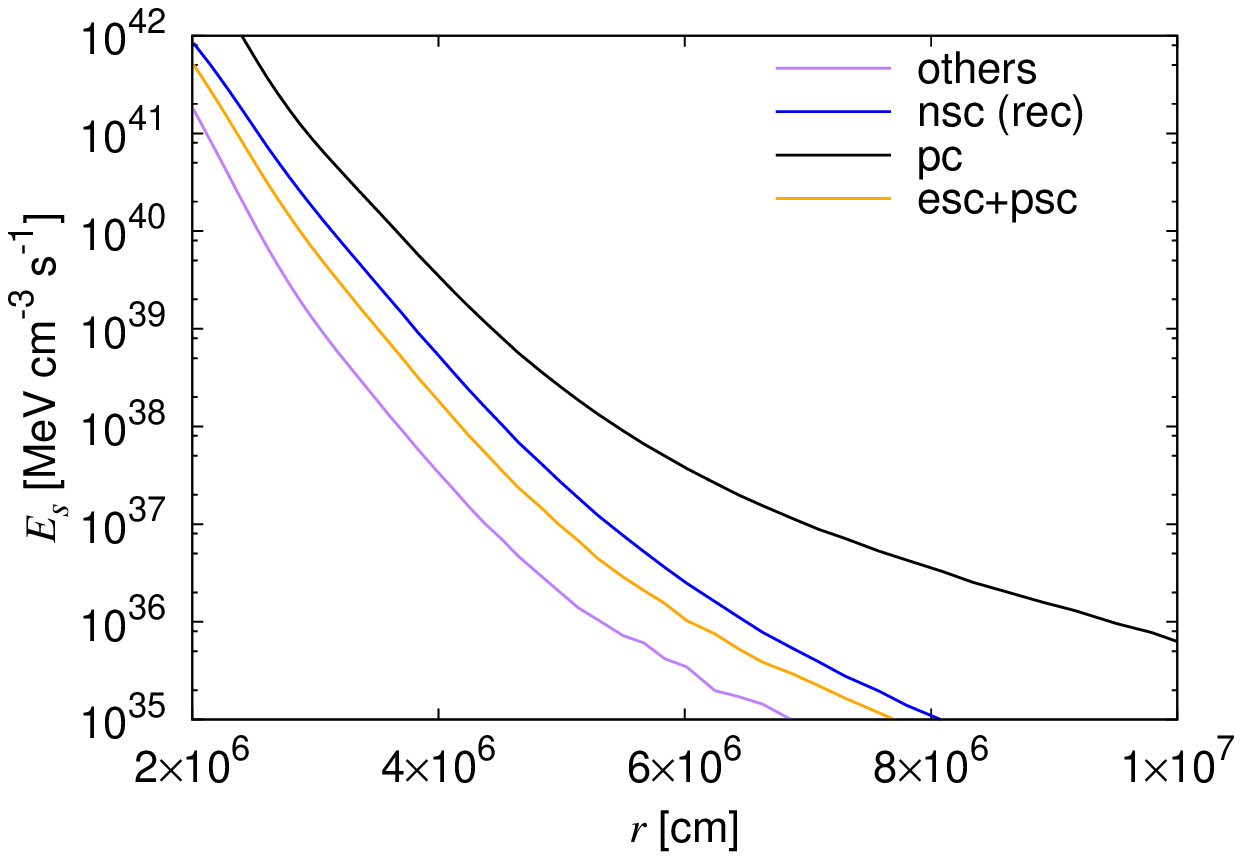}
\plotone{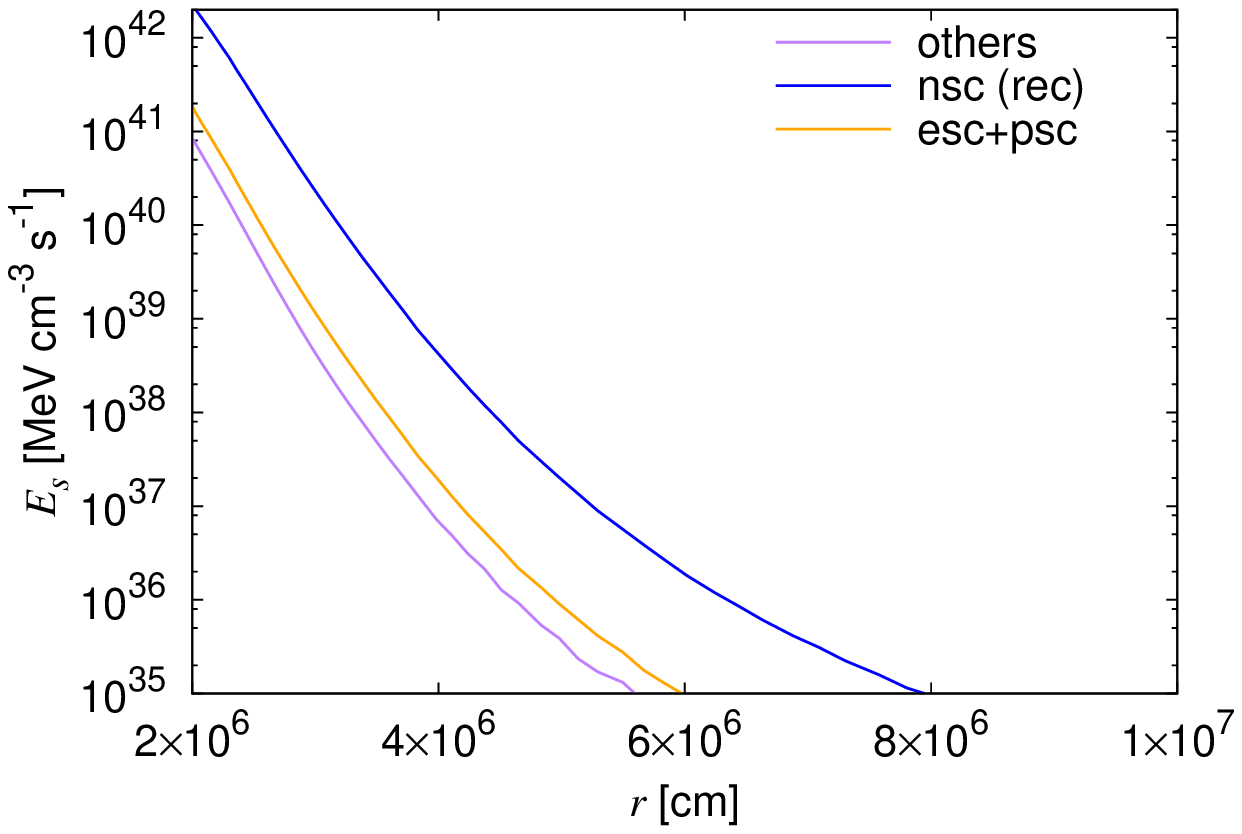}
\caption{The same as the right panels of Figure~\ref{reac:rec} except for the inclusion of the electron/positron scattering (orange). \label{reac:esc}}
\end{figure}

Figure~\ref{reac:esc} is the same as the right panels of Figure~\ref{reac:rec} except for the addition of the electron/positron scattering as shown in orange.
We find that apart from the charged-current reactions for $\nu_e$'s and $\bar{\nu}_e$'s, the accumulation of small recoils in the nucleon scattering is more important than a smaller number of large recoils in the electron/positron scattering in the thermalization of neutrinos at least for this particular model.
Indeed, we find that the energy spectra of neutrinos are almost identical to those without the electron/positron scattering\footnote{Note that the cross section of the electron/positron scattering for low energy neutrinos ($\sim$ a few MeV) is higher than that of the nucleon scattering. On the other hand, those low energy neutrinos have already decoupled from matter, and hence the energy spectra of neutrinos at low energy are less sensitive to the change of the cross sections.} (see Figure~\ref{spe:rec}).
%The energy spectra are not changed by the inclusion of the electron/positron scattering as we expect easily.
%Although the electron/positron scattering has a larger cross section than the nucleron scattering at $E_\nu \lesssim$ a few MeV, the region of interest is optically thin to neutrinos with such a low energy. 
Note also that \cite{2000PhRvC..62c5802T} calculated the thermalization of $\nu_x$'s in a uniform background matter with their own MC code and reached the same conclusion.

%%%%%%%%%%%%%%%%%%%%%%%%%%%%%%%%%%%%%%%%%%%%%%%%%%%%%%%%%%%%%%%%%%%%%%%%%%%%%%%%%

\section{Implications for the numerical implementation of nucleon recoils in the finite-difference method} \label{ch5}

The nucleon recoil affects the neutrino luminosity and dynamics of explosions as discussed in the literature \citep{2002A&A...396..361R,2006A&A...447.1049B,2009ApJ...694..664M,2010PhRvL.104y1101H,2012ApJ...760...94L,2012ApJ...761...72M,2015ApJ...808..188P,2015ApJ...807L..31L,Skinner:2015uhw,2017ApJ...850...43R,2018ApJ...853..170K,2018arXiv180905608B,2019MNRAS.482..351V,2019MNRAS.485.3153B,2019arXiv190110523R,2019ApJ...873...45G}.
Although the finite-difference method is normally employed for neutrino transport in the CCSNe simulations, we can not afford to deploy a sufficiently large number of energy bins needed to resolve the small energy exchange via the nucleon recoil in each scattering.
Some sub-grid technique is hence adopted \citep{2006A&A...447.1049B}.
In this section we conduct some experimental MC runs to investigate possible consequences of such numerical implementations of the nucleon recoil in the finite-difference transport schemes such as the $S_N$ method.
We quantify the effects of coarse-energy grids on the energy spectrum of neutrino and present a possible improvement.

When the cell width of the energy grid is much larger than the typical value of the energy exchange in the scattering, it is certainly inappropriate to use the cell-center values of energies and neutrino distribution functions to evaluate the rate of the scattering that transfer neutrinos in one energy cell to another adjacent to it.
This is because those neutrinos existing in the close vicinity of the energy-cell boundary can cross it over to the next cell.
In the finite-difference method adopting such an energy grid, it is required to reconstruct the neutrino distribution inside the energy bin somehow to estimate the neutrino populations near the cell boundary and calculate the scattering rate based on them; once the neutrinos enter the next energy cell, they are mixed with others in the same cell and their individual energies are forgotten.
We mimic such a situation in the MC simulation by introducing the energy grid and re-distributing MC particles in each energy bin after a certain interval in a couple of ways and study how the results are affected.
% \citep{Sumiyoshi:2012za,2009ApJ...698.1174L,2018ApJ...854...63O,1995ApJ...450..830B}.
%The widths of bins are, however, finite and we have to impose the artificial distribution in each bins, which may produce 

We adopt three artificial ways of the re-distribution in each energy bin: ``flat'', ``linear+Ncons'' and ``linear+NEcons''.
The first one is the simplest but the coarsest reconstruction, in which we homogenize the distribution of sample particles in each energy bin.
In the second and third cases we introduce linear distributions.
The inclination and intercept of the linear functions are determined in both cases so that the number of the MC particles should be unchanged and in the second case the values at the two neighbor cells are employed in the interpolation.
In the third case, on the other hand, we impose the energy conservation in the reconstruction.
The distribution of sample particles in the $k$-th energy bin is given as follows:
\begin{eqnarray}
\frac{dN_{T,k}}{dE} = a_kE + b_k,
%\Delta N_{\nu,k} = \left(aE_{\nu,k} + b \right)\Delta E_{\nu,k} ,
\end{eqnarray}
with the inclination, the intercept and the total number of sample particles in the $k$-th energy bin $a_k$, $b_k$ and $N_{T,k}$, respectively.
We determine $a_k$ by the weighted average of two inclinations $a_1$ and $a_2$,
\begin{eqnarray}
&& a_k = a_1\frac{E_{\nu,k+1}-E_{\nu m,k}}{E_{\nu,k}-E_{\nu,k-1}} 
     + a_2 \frac{E_{\nu m,k}-E_{\nu,k-1}}{E_{\nu,k}-E_{\nu,k-1}}, \label{a_k} \\ 
&& a_1 = \frac{N_{T,k}/(E_{\nu,k+1}-E_{\nu,k})}{E_{\nu m,k+1} - E_{\nu m,k}}, \\
&& a_2 = \frac{N_{T,k}/(E_{\nu,k}-E_{\nu,k-1})}{E_{\nu m,k} - E_{\nu m,k-1}}, \ \ \ \ 
\end{eqnarray}
with the mid-point energy of the $k$-th energy bin $E_{\nu m,k}$ and obtain $b_k$ from solving the equation for $N_{T,k}$,
\begin{eqnarray}
N_{T,k} = \int_{E_{\nu,k-1}}^{E_{\nu,k}} \left(a_kE + b_k\right)\ dE, \label{N_TK}
%b = \frac{ N_{T,k}-\frac{1}{2}a \left(E_{\nu,k}^2-E_{\nu,k-1}^2\right) }{ E_{\nu,k}-E_{\nu,k-1}},
\end{eqnarray}
for the second case, while we adopt eq.~(\ref{N_TK}) and the equation for the total energy of the $k$-th energy bin $E_{T,k}$,
\begin{eqnarray}
E_{T,k} = \int_{E_{\nu,k-1}}^{E_{\nu,k}} \left(a_kE + b_k\right)E\ dE,
%&&a \equiv - \frac{12}{\left(E_{\nu,k}-E_{\nu,k-1}\right)^3} \nonumber \\
%&& \ \ \ \ \ \ \ \ \times\left[\frac{1}{2}N_{T,k}\left(E_{\nu,k-1}+E_{\nu,k}\right)-E_{T,k}\right], \\ 
%&&b \equiv \frac{4N_{T,k}\left(E_{i}^3 - E_{\nu,k-1}^3\right)-6E_{T,k}\left(E_{i}^2-E_{\nu,k-1}^2\right)}{\left(E_{i}-E_{\nu,k-1}\right)^4},\ \ \ \ \ \ 
\end{eqnarray}
to determining $a_k$ and $b_k$ for the third case.

We introduce two energy grids with different numbers of cells: $N_{E_\nu}$ = 10 and 20 to cover the energy range of 0 -- 300 MeV.
Note that the latter is exactly the same as the energy grid employed in the Boltzmann solver by \cite{2019ApJS..240...38N}.
We focus on the spectra of $\nu_x$'s, which are affected most by the inclusion of nucleon recoils as shown in the previous sections. 
The artificial re-distributions of sample particles in each energy bin are repeated on the time scale of a single time step of CCSN simulations to mimic their situation in the finite-difference method.
We adopt as a background the same hydrodynamical model as that employed in the previous sections (see Figure~\ref{hydro}) and deploy the same number of sample particles and use the same $dt_f$ as well.
We run the MC code for the spectrum obtained in the previous steady-state calculations with the re-distribution implemented.
The r1 set of neutrino reactions is adopted.
We take the average of the distribution function over 8,000 time steps after the steady-state is achieved.

\begin{figure}[htbp]
\epsscale{1.2}
\plotone{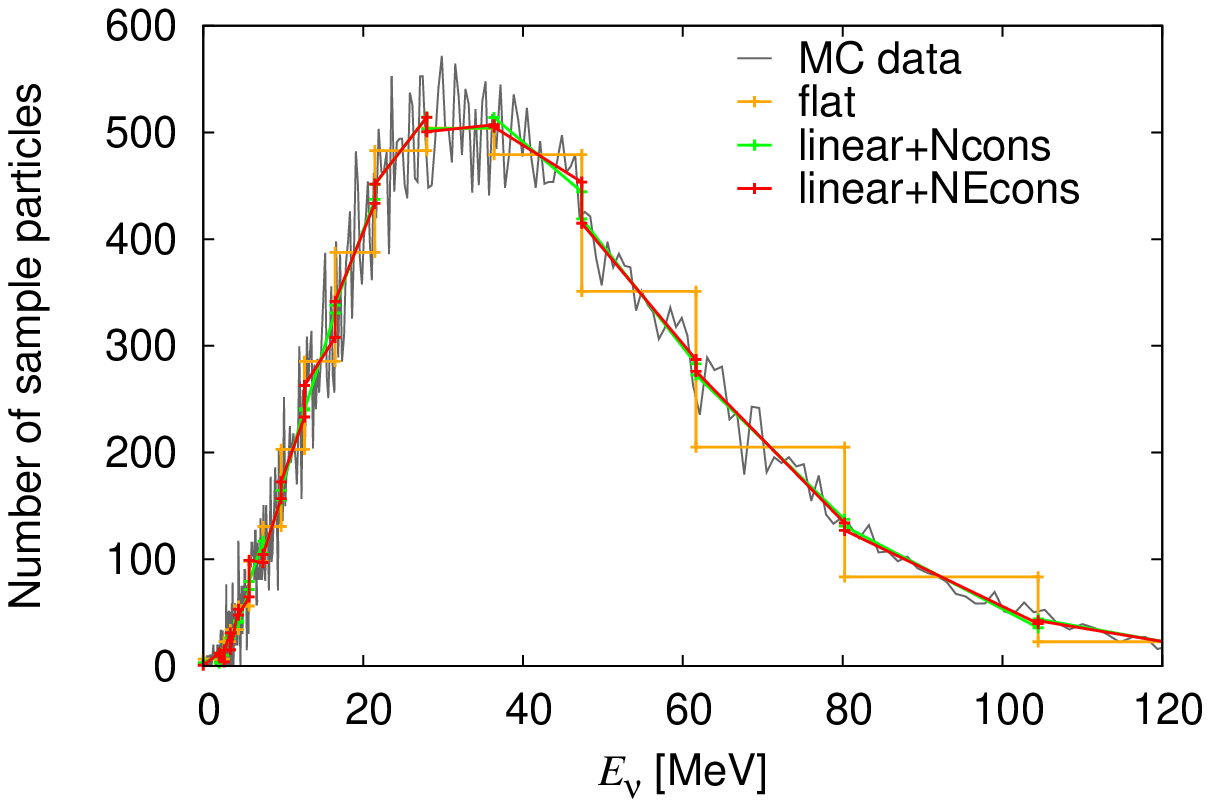}
\plotone{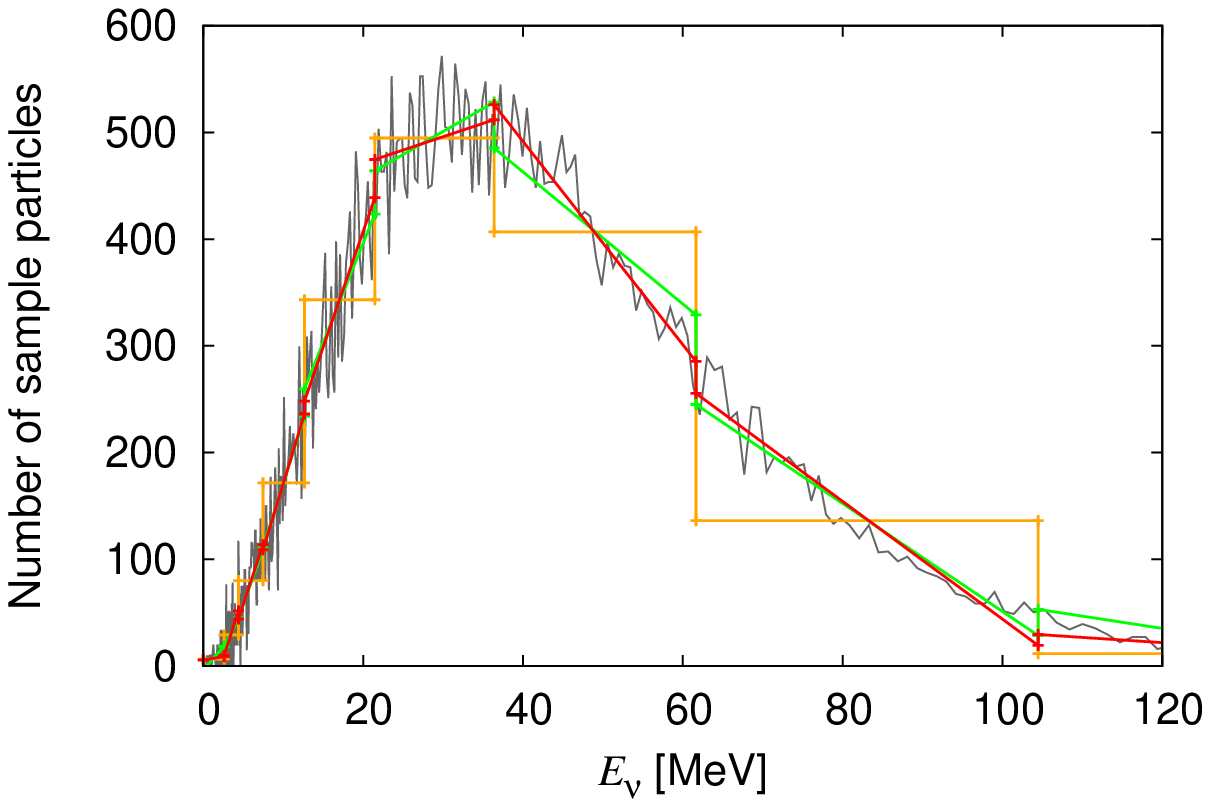}
%\plotone{check_inc_conserv.eps}
\caption{The energy spectra of $\nu_x$'s for $N_{E_\nu}$ = 20 (top) and 10 (bottom). The gray line is the original spectrum obtained without re-distribution, whereas the other lines correspond to the different re-distribution models: flat (orange), linear+Ncons (green) and linear+NEcons (red). \label{check_inc}}
\end{figure}

Figure~\ref{check_inc} demonstrates the three different reconstructions of neutrino spectrum described above for the two energy grids with $N_{E_\nu}$ = 20 (top panel) and 10 (bottom panel).
The gray line is the original spectrum obtained by the MC calculation without re-distribution.
The lines with other colors denote the spectra reconstructed as explained above.
In the case of $N_{E_\nu}$ = 20 the linear+Ncons and linear+NEcons models give similar distributions (see the green and red lines), whereas they are more deviated from each other for $N_{E_\nu}$ = 10.
This difference turns out to be an important later.
%The models except for the flat model seem to repdroduce the original spectrum well in the case of $N_{E_\nu}$ = 20, whereas the spectra reconstructed by the linear and linear+Ncons models differ from the original one, especially at high energies in the case of $N_{E_\nu}$ = 10.

\begin{figure*}[htbp]
\epsscale{1.0}
\plottwo{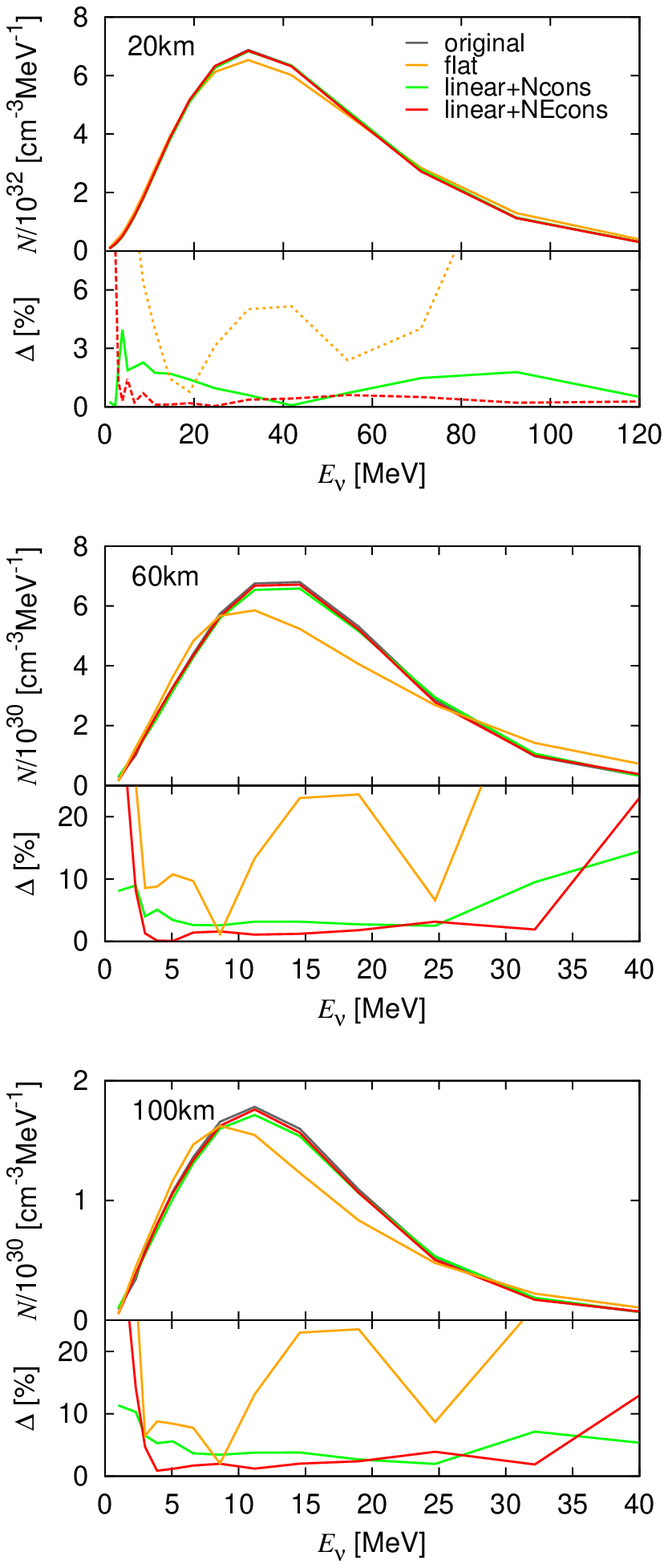}{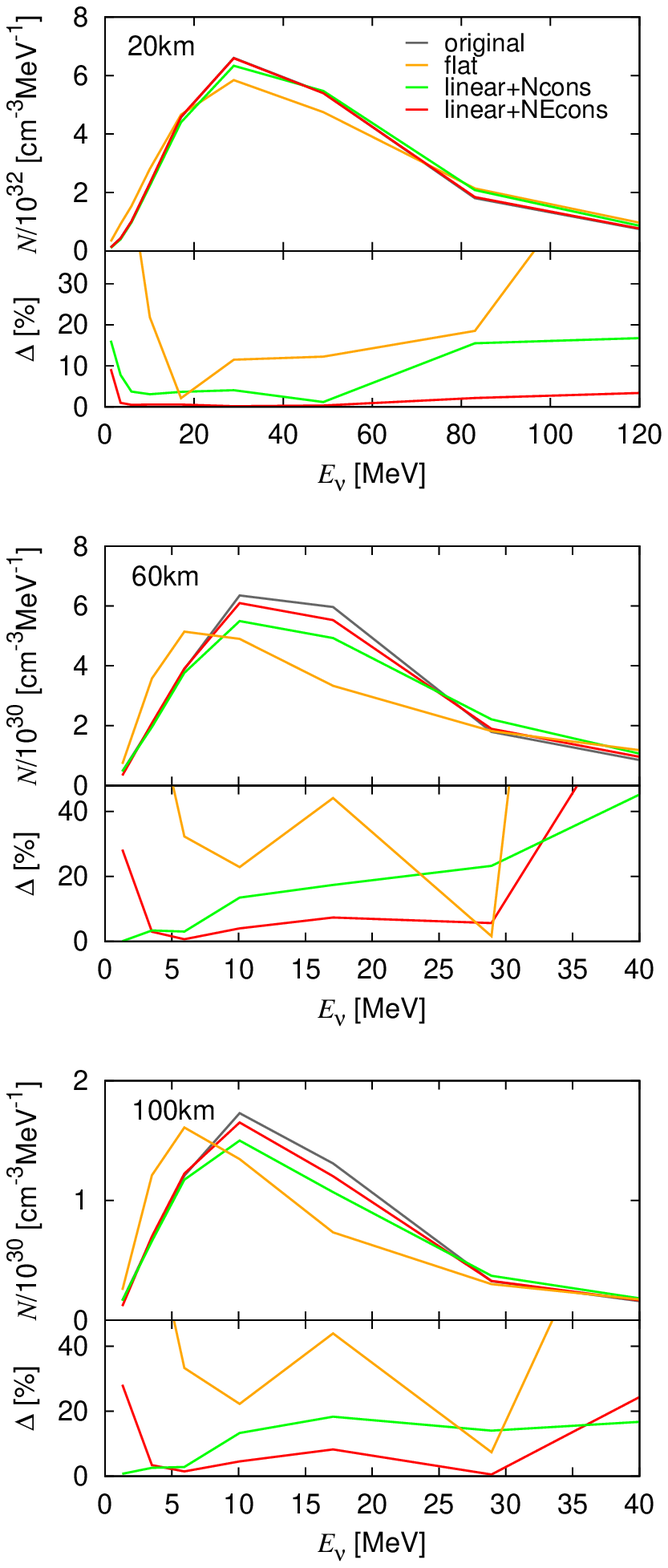}
%\plottwo{compair_initial_flat.eps}{compair_initial_conserv.eps}
\caption{The energy spectra of $\nu_x$'s at $r$ = 20 (top), 60 (middle) and 100 km (bottom) obtained with three artificial neutrino re-distributions: flat (orange), linear+Ncons (green) and linear+NEcons (red). The left and right panels show the results with the number of grids $N_{E_\nu}$ = 20 and 10, respectively. The gray lines denote the original, supposedly correct neutrino spectra derived from the previous steady-state calculations without re-distribution. The relative errors $\Delta$ are shown for the same models in the lower half panels. \label{compair_initial}}
\end{figure*}

Figure~\ref{compair_initial} shows the resultant state distributions (upper half) and the deviations from the original, supposedly correct ones $\Delta$ (lower half) at $r$ = 20 (top), 60 (middle) and 100 km (bottom).
The color coding is the same as before.
%Gray lines show the correct spectra and the other color lines denote the ones derived from the steady-state calculations with four artificail distribution models.
In the case of $N_{E_\nu}$ = 20 presented in the left panels, we find that the flat re-distribution produces errors as large as $\sim$20\% near the average energy (see the orange lines).
This is because a larger number of sample particles can get across the boundaries of energy bins and move to the next cells thanks to the re-distribution and may be regarded as the overestimation of the energy exchange via nucleon recoils.

In the two linear re-distribution models (the green and red lines) the error is reduced to a few\%.
We find smaller differences in the former model at lower and higher energies, whereas the latter model reproduces the peak of neutrino spectra better.
It is difficult to say which of the two is better from these results.
If we reduce the number of energy grids to $N_{E_\nu}$ = 10 (right panels), however, their results differ more from each other.
The error $\Delta$ in the liner+NEcons model increases but still stay within 10\% for almost all energies even at large radii.
The spectra for the linear+Ncons model, on the other hand, deviate from the correct ones by $\sim$ 20\%.
This difference is a consequence of the difference in the re-distributions, which we found becomes remarkable when the energy grid gets coarser.
Note that $N_{E_\nu}$ = 10 is not very low compared to that employed in current CCSNe simulations.
We had better hence impose, if possible, the energy conservation in reconstructing the neutrino distribution in each energy bin to incorporate the effects of nucleon recoils in neutrino transport accurately, particularly when the energy resolution is not high.
This will be possible if not only the number but also the energy in each energy bin is stored in the transport.

\section{Summary and discussions} \label{ch6}

The nucleon recoils in the neutrino-nucleon scattering is one of the important factors for the dynamics of supernova explosions and neutrino observations and their effects have been already investigated in the literature.
In these studies the finite-difference method is normally adopted for neutrino transport.
In so doing, we cannot afford to deploy a sufficiently large number of energy bins needed to resolve the small energy exchange in the nucleon recoil.
In this paper we have performed neutrino transport calculations with our own MC code for a static matter background derived from a dynamical SN simulation to quantify the effects of the coarse energy grid and suggest a possible improvement in the sub-grid modeling.

%%code validation
We have first conducted two test calculations for the validation of our MC code.
We have compared steady-state solutions obtained with the MC code and those with our finite-difference Boltzmann solver, in which we employ a matter background computed from one of our recent CCSN simulations \citep{2019ApJS..240...38N}.
The nucleon recoil has been ignored in this comparison.
We have demonstrated that the two results are in excellent agreement with each other.
In order to confirm the detailed balance in our treatment of the nucleon recoil, we have done a one-zone calculation of the thermalization of neutrino spectrum via the neutron scattering.
This is ensured by calculating the reaction rates only for $E_\nu \le E^\prime_\nu$ and deriving those for $E_\nu > E^\prime_\nu$ from them via the detailed balance relation.
We have confirmed indeed that the neutrino spectrum approaches a thermal distribution as expected.

%%%nucleon recoils effect
We have then run the MC code to compute the thermalization of energy spectra as neutrinos propagate outwards in the post-shock region.
We have first studied the large proton mass limit of the proton scattering, in which it becomes iso-energetic, and have made clear three important effects of the recoil on its reaction rate: the broadening of neutrino spectra, the reduction of the cross section and the change of the angle dependence of the reaction rate.
We have then re-applied the MC code to the neutrino transport calculations on the same static matter background as that employed in the code validation but with the nucleon recoil being incorporated this time.

We have found a significant change in the spectra of $\nu_x$'s by the inclusion of the nucleon recoil.
High-energy $\nu_x$'s are depleted while low-energy ones are increased due to down-scatterings and their average energy is reduced by $\sim$15\%.
The spectra of $\nu_e$'s and $\bar{\nu}_e$'s, on the other hand, do not change much by the inclusion of the nucleon recoil.

These different responses to the nucleon recoil among different flavors of neutrinos are explained as follows.
The number of nucleon scatterings is smaller than that of electron captures by a factor $\sim$ 5 for $\nu_e$'s, whereas the dominant reaction for $\nu_x$'s is the nucleon scattering.
For $\bar{\nu}_e$'s the number of the nucleon scattering is larger than those of other reactions, which seems to contradict with the result that the spectrum of $\bar{\nu}_e$'s is not changed by the nucleon recoil.
The reason is simply because the energy exchange in the nucleon scattering is much smaller.

Next, we have incorporated the electron/positron scattering in the MC code and compared the contributions to thermalization between the two scatterings.
The energy exchange per scattering for the electron/positron scattering is much larger than that for the nucleon scattering because of the smaller mass of the former, $m_e = 0.511$ MeV, whereas the cross section of the latter is larger than that of the former at $E_\nu \gtrsim$ a few MeV.
We have found that the accumulation of small recoils in the nucleon scattering is more important than a smaller number of large recoils in the electron/positron scattering in the thermalization of neutrinos at least for this particular model.

%%implementation
We have then conducted some experimental MC runs to investigate the implications for the numerical implementation of the nucleon recoil in the finite-difference transport schemes, which have been frequently employed in CCSN simulations.
The width of energy bins employed in these schemes is normally much larger than the typical energy exchanged via the nucleon recoil and the sub-grid modeling is somehow needed.
In order to mimic such situations, we have introduced energy grids in the experimental MC runs and artificially re-distributed sample particles repeatedly after a typical time interval in the CCSNe simulations.
We have considered three artificial distributions of neutrinos in each energy bin and referred to them as ``flat'', ``linear+Ncons'' and ``linear+NEcons''.
In this study, we have adopted two energy grids with different numbers of grid points: $N_{E_\nu}$ = 10 and 20.
Note that the latter grid is exactly the same as that employed in our axisymmetric CCSN simulations with the Boltzmann solver \citep{nagakura2018,2019ApJ...880L..28N}.
We have run the MC code with this re-distribution scheme implemented for the same matter background as that in the previous calculations without re-distributions.
We have found that the neutrino spectra in the flat model are deviated from the correct one by $\sim$20\% even in the high energy resolution $N_{E_\nu}$ = 20, whereas the difference is reduced to a few \% in the linear+Ncons and linear+NEcons models.
Both of the latter two models can reconstruct the original spectra equally accurately for $N_{E_\nu}$ = 20.
If we reduce the number of energy grid points to $N_{E_\nu}$ = 10, however, their results differ from each other.
Although the errors in the liner+NEcons model are still within 10\% at almost all energies even in the outer region, they rise up to $\sim$ 20\% in the linear+Ncons model.
Since the energy resolution typically employed in the finite-difference methods is rarely higher than the $N_{E_\nu}$ = 20 case in this paper, it is recommended to keep the track of not only the number but also the energy in each energy bin somehow and use the number and energy conservations to reconstruct the sub-grid distributions of neutrinos when dealing with the small energy exchange in the nucleon recoil.

%Future work
%%Boltzmann problem
Our next task is to actually implement these sub-grid modellings into the Boltzmann solver, in which they are employing the reaction rate of \cite{1985ApJS...58..771B} for the nucleon scattering currently, and to perform CCSNe simulations.
This will enable us to discuss the effects of the nucleon recoil, particularly its energy-resolution dependence, on the dynamics of explosion and PNS cooling quantitatively.
It should be also important from the observational point of views.

\acknowledgments{This work was partly supported by Research Fellowships of Japan Society for the Promotion of Science (JSPS). H.N. was supported by Princeton University through DOE SciDAC4 Grant DE-SC0018297 (subaward 00009650). Numerical computations were carried out on Cray XC50 at Center for Computational Astrophysics, National Astronomical Observatory of Japan.} 

\bibliography{paper}


%\appendix
\appendix

\section{Determination of neutrino energy after scattering} \label{appendix}

In our code, we employ the reaction-rate tables for the nucleon and electron/positron scatterings.
In order to ensure the detailed balance between the direct and inverse reactions between the initial and final states with the neutrino energies $E_\nu$ and $E^\prime_\nu$, respectively, we take the following method.

\begin{enumerate}
\item{$E_\nu \le E^\prime_\nu \le E_{\rm{max}}$}\\
The reaction rates for \textcolor{black}{up-scatterings} $E_\nu \le E^\prime_\nu$ are included in the table and we get $E^\prime_\nu$ interpolating data in the table.
We use the modified reaction rate $\bar{R}$ instead of $R_{\rm{rec}}$ for convenience:
\begin{eqnarray}
&&\bar{R}\left(E_\nu,\Delta E,\cos{\psi}\right) = R_{\rm{rec}}\left(E_\nu,E^\prime_\nu,\cos{\psi}\right)\exp{\left(-\frac{E_\nu}{T}\right)},
\end{eqnarray}
with the energy difference $\Delta E \equiv E^\prime_\nu-E_\nu$.
The modified reaction rates are described by the reaction rates in the table $\bar{R}_{ij} \equiv \bar{R}\left(E_i,\Delta E_{ij},\cos{\psi}\right)$ with the on-grid neutrino energy $E_i$ employed in the table $E_1 \le E_\nu \le E_2$ and $E^\prime_1 \le E^\prime_\nu \le E^\prime_2$ and the energy difference $\Delta E_{ij} \equiv E^\prime_j - E_i$:
\begin{eqnarray}
\bar{R}\left(E_\nu,\Delta E,\cos{\psi}\right) = q_1k_1\bar{R}_{11} + q_1k_2\bar{R}_{12} + q_2k^\prime_1\bar{R}_{21} + q_2k^\prime_2\bar{R}_{22},
\end{eqnarray}
where the coefficients are defined as follows:
\begin{eqnarray}
&&q_1 = \frac{E_2-E_\nu}{E_2-E_1},\ \ q_2 = \frac{E_\nu-E_1}{E_2-E_1}, \\
&&k_1 = \frac{\Delta E_{12} - \Delta E}{\Delta E_{12} - \Delta E_{11}},\ \ k_2 = \frac{\Delta E - \Delta E_{11}}{\Delta E_{12} - \Delta E_{11}}, \\
&&k^\prime_1 = \frac{\Delta E_{22} - \Delta E}{\Delta E_{22} - \Delta E_{21}},\ \ k^\prime_2 = \frac{\Delta E - \Delta E_{21}}{\Delta E_{22} - \Delta E_{21}}.
\end{eqnarray}

\item{$E_{\rm{min}} \le E^\prime_\nu \le E_\nu$}\\
The reaction rates for \textcolor{black}{down-scatterings} $E_\nu \ge E^\prime_\nu$ are derived from the rates for \textcolor{black}{up-scatterings} $E_\nu \le E^\prime_\nu$ using the following relation:
\begin{eqnarray}
\bar{R}\left(E_\nu,E^\prime_\nu,\cos{\psi}\right) = \bar{R}\left(E^\prime_\nu,E_\nu,\cos{\psi}\right),
\end{eqnarray}
based on the detailed balance.
The modified reaction rate is described as
\begin{eqnarray}
&&\bar{R}\left(E^\prime_\nu,E_\nu,\cos{\psi}\right) = q_3k_3\bar{R}_{33} + q_3k_4\bar{R}_{34} + q_4k^\prime_3\bar{R}_{43} + q_4k^\prime_4\bar{R}_{44}.
\end{eqnarray}
with the neutrino energy employed in the table $E_3 \le E^\prime_\nu \le E_4$ and $E^\prime_3 \le E_\nu \le E^\prime_4$, the energy difference $\Delta E^\prime \equiv E_\nu - E^\prime_\nu$ and the coefficients:
\begin{eqnarray}
&&q_3 = \frac{E_4-E^\prime_\nu}{E_4-E_3},\ \ q_4 = \frac{E^\prime_\nu-E_3}{E_4-E_3}, \\
&&k_3 = \frac{\Delta E_{34} - \Delta E^\prime}{\Delta E_{34} - \Delta E_{33}},\ \ k_4 = \frac{\Delta E^\prime - \Delta E_{33}}{\Delta E_{34} - \Delta E_{33}}, \\
&&k^\prime_3 = \frac{\Delta E_{44} - \Delta E^\prime}{\Delta E_{44} - \Delta E_{43}},\ \ k^\prime_4 = \frac{\Delta E^\prime - \Delta E_{43}}{\Delta E_{44} - \Delta E_{43}},   
\end{eqnarray}
\end{enumerate}

The total rate integrated over $E^\prime_\nu$ is
\begin{eqnarray}
A &\equiv& \int^{E_{\rm{max}}}_{E_{\rm{min}}} R \left(E_\nu,\bar{E}_\nu,\cos{\psi}\right) 2\pi \bar{E}^2_\nu d\bar{E}_\nu \nonumber \\
&=& \int^{E_{\rm{max}}}_{E_{\rm{min}}} \bar{R}\left(E_\nu,\bar{E}_\nu,\cos{\psi}\right)\exp{\left(\frac{E_\nu}{T}\right)}2\pi \bar{E}^2_\nu d\bar{E}_\nu \nonumber \\
&=& 2\pi \exp{\left(\frac{E_\nu}{T}\right)} \left[ \int^{E_\nu}_{E_{\rm{min}}} \bar{R}\left(E_\nu,\bar{E}_\nu,\cos{\psi}\right) \bar{E}^2_\nu d\bar{E}_\nu + \int^{E_{\rm{max}}}_{E_\nu} \bar{R}\left(\bar{E}_\nu,E_\nu,\cos{\psi}\right) \bar{E}^2_\nu d\bar{E}_\nu \right] \nonumber \\
&=& \frac{1}{4}\left(E^4_\nu-E^4_{\rm{min}}\right)A_{11} + \frac{1}{3} \left(E^3_\nu - E^3_{\rm{min}}\right)A_{12} \nonumber \\
&& \ + \frac{1}{5} \left(E^5_{\rm{max}}-E^5_\nu\right)B_{11} + \frac{1}{4} \left(E^4_{\rm{max}}-E^4_\nu\right)B_{12}+ \frac{1}{3} \left(E^3_{\rm{max}}-E^3_\nu\right)B_{13}, 
\end{eqnarray}
with the minimum and maximum energies $E_{\rm{min}}, E_{\rm{max}}$, at which the reaction rates become $10^{-5}$ times less than the peak value, and the coefficients:
\begin{eqnarray}
A_{11} &=& \frac{- \bar{R}_{11} + \bar{R}_{12}}{\Delta E_{12}-\Delta E_{11}}q_1 + \frac{- \bar{R}_{21} + \bar{R}_{22}}{\Delta E_{22} - \Delta E_{21}}q_2, \\
A_{12} &=& \frac{\left(\Delta E_{12} + E_\nu\right)\bar{R}_{11} - \left(\Delta E_{11} + E_\nu\right)\bar{R}_{12}}{\Delta E_{12}-\Delta E_{11}}q_1 \nonumber \\
&&\ \ \ \ \        + \frac{\left(\Delta E_{22} + E_\nu \right) \bar{R}_{21} - \left(\Delta E_{21} + E_\nu\right) \bar{R}_{22}}{\Delta E_{22} - \Delta E_{21}}q_2, \\
B_{11} &=& \frac{1}{E_4 - E_3} \left(\frac{-\bar{R}_{33}+\bar{R}_{34}}{\Delta E_{34} - \Delta E_{33}} + \frac{\bar{R}_{43}-\bar{R}_{44}}{\Delta E_{44} - \Delta E_{43}}\right),\\
B_{12} &=& \frac{1}{E_4 - E_3} \left(\frac{\bar{R}_{33}\left(E_4-\Delta E_{34} + E_\nu\right) - \bar{R}_{34} \left(E_4-\Delta E_{33} + E_\nu\right)}{\Delta E_{34} - \Delta E_{33}} \right. \nonumber \\
&&\ \ \ \ \ \ \ \ \ \ \ \ \ \ \ \left. + \frac{\bar{R}_{43} \left(\Delta E_{44} - E_\nu -E_3\right) -\bar{R}_{44}\left(\Delta E_{43}-E_\nu-E_3\right)}{\Delta E_{44} - \Delta E_{43}}\right),\\
B_{13} &=& \frac{1}{E_4 - E_3} \left(\frac{\bar{R}_{33}E_4\left(\Delta E_{34} - E_\nu\right) + \bar{R}_{34} E_4\left(E_\nu - \Delta E_{33}\right)}{\Delta E_{34} - \Delta E_{33}} \right. \nonumber \\
&& \ \ \ \ \ \ \ \ \ \ \ \ \ \left. + \frac{\bar{R}_{43} E_3\left(E_\nu - \Delta E_{44}\right) + \bar{R}_{44}E_3\left(\Delta E_{43}-E_\nu\right)}{\Delta E_{44} - \Delta E_{43}}\right).
\end{eqnarray}
The neutrino energy after scattering $E^\prime_\nu$ is determined by the random number $x$ in the range of [0,1] and the normalized spectrum $\int^{E^\prime_\nu}_{E_{\rm{min}}}R \left(E_\nu,\bar{E}_\nu,\cos{\psi}\right) 2\pi \bar{E}^2_\nu d\bar{E}_\nu/A$.

\section{Numerical method of our MC code} \label{appendix2}

\subsection{Sample particles} \label{subch:MC}
%%%%%%idea of MC study
In the MC method, we follow the tracks of sample particles, which represent a bundle of neutrinos, interacting with matters.
%%%%%decision of weight
The numbers of sample particles $N_{s}$ and physical neutrinos $N_{\nu}$ are related with the weight $W_{s}$ as follows:
\begin{eqnarray}
W_{s} = \frac{ N_{\nu} }{ N_{s} } .
\end{eqnarray}
In our simulations, the weight is constant in all the time and calculation domain.

\subsection{Treatments of the transport of sample particles} \label{subch:trans}
%%%%%%propagation of monte Carlo sample
Each sample particle has 6-dimensional information about a space $(r,\theta,\phi)$ and a phase space $(E_{\nu},\theta_{\nu},\phi_{\nu})$, and we calculate their time evolutions by solving geometric equations.
In order to calculate the transport of sample particles, we introduce three lengths : ``reaction length'' $l_{\text{r}}$, ``background length'' $l_{\text{b}}$ and ``distribution length'' $l_{\text{f}}$.

\begin{enumerate}
\item{reaction length $l_{\text{r}}$ \\}
We define a ``reaction length'', which is a distance to the point where the sample particle interacts with matter subsequently, by the optical depth:
\begin{eqnarray}
\tau(S,E_\nu) = \int_0^{S} \frac{1}{\lambda(r,E_\nu)} ds ,
\end{eqnarray}
using the local mean free path $\lambda$:
\begin{eqnarray}
\lambda(r,E_\nu) = \frac{1}{\sigma_{\text{tot}}},
\end{eqnarray}
with the total cross section $\sigma_{\text{tot}} = \sum_{\alpha}^{}\sigma_\alpha(r,E_\nu)$ using the cross section of $\alpha$-th type of reaction $\sigma_\alpha$.
The reaction occurs at $\tau(l_{\text{r}},E_\nu) = \tau_{\text{max}}$, which is determined by the random number obeying the Poisson distribution whose average becomes 1.

\item{background length $l_{\text{b}}$ \\}
We employ the results of the dynamical SN simulations as a background for the neutrino transport calculations.
\textcolor{black}{We assume that the hydrodynamical values, i.e. density, temperature and chemical potential of matters, are uniform in each spatial zone.}
A ``background length'' is defined by the distance between the nearest spatial boundary of the hydrodynamical background and the current position of a sample particle.

\item{distribution length $l_{\text{f}}$\\}
The distribution functions of neutrinos change with time because of interactions with matter or advection. 
We have to update it within an appropriate timescale, because the Fermi-blocking of neutrinos should be taken into account for neutrino reactions. 
A "distribution length'' is defined as $cdt_{\rm{f}}$ with the remaining time until the update of the distribution function $dt_{\rm{f}}$ (``distribution time'').

\end{enumerate}

Sample particles can propagate independently, but their global times have to be coincident updating the neutrino distribution function. 
We hence take a time step of calculations as the distribution time $dt = dt_{\text{f}}$ and calculate the individual evolution of the sample particle during each time step.
If the other two lengths are longer than the distribution length, the sample particle of interest just propagates freely during this time step.
If not, comparing between the reaction and background lengths, this sample undergoes the process with the shorter length, subsequently, and we recalculate these lengths.
We repeat this cycle for each sample particle until the distribution time $dt_{\rm{f}}$ elapses. 
After calculating the evolutions of all sample particles, individually, we update the distribution function as described in Section~\ref{subch:f}.

%The algorithm of the transport of sample particles is summarized in Fig-.

\subsection{\textcolor{black}{Evaluation} of the neutrino distribution function} \label{subch:f}
%%%%%%making f
In this calculation, we employ the spherical symmetric background and the neutrino distribution function is reduced to $f(r,E_{\nu},\theta_{\nu})$.
At every time step, we count the number of sample particles inside each volume element in a space and a phase space, and calculate the $i,j,k$-th discretized neutrino distribution function $f_{ijk}$:
\begin{eqnarray}
f_{ijk} = \frac{N_{ijk}W_{s}}{\textcolor{black}{V_{r,i}V_{m,jk}}},
\end{eqnarray}
where $i, j$ and $k$ describe the components of $r,\ E_\nu$ and $\theta_\nu$, respectively; the total number of sample particles in the $i,j,k$-th volume element $N_{ijk}$; the $i$-th spatial volume element $\textcolor{black}{V_{r,i}} = 4\pi\left(r^3_i-r^3_{i-1}\right)/3$ and the $j,k$-th phase space volume element $\textcolor{black}{V_{m,jk}} = 2\pi\left(\cos{\theta_{\nu,k}}-\cos{\theta_{\nu,k-1}}\right)\left(E^3_{\nu,j}-E^3_{\nu,j-1}\right)/3$.

\subsection{Treatments of neutrino reactions} \label{subch:2.5}
Neutrinos interact with matter via several reactions inside stars (See Table~\ref{reac_MC}).
We divide neutrino reactions into three processes: absorption, emission and scattering, and adopt different treatments to them in our MC code.

\subsubsection{Absorption and scattering}
Existing samples are absorbed or scattered by matter.
After the subsequent reaction point is determined by the reaction length, which is defined by the mean free path of all absorption and scattering processes taken into account, we choose which reaction will occur actually using the uniform random number $x$ whose range is [0, 1]. 
If we get the random number in the range of $\Sigma_{\alpha=1}^{i-1} \sigma_\alpha/\sigma_{\text{tot}} \leqq x < \Sigma_{\alpha=1}^{i} \sigma_{\alpha}/\sigma_{\text{tot}}$, the sample particle will undergo the $i$-th reaction \citep{1978ApJS...37..287T,Lucy:2003zx}.
If the $i$-th reaction is an absorption process, such as $\nu_e + n \rightarrow p + e^-$, we stop following the track of this sample particle at this point.
If the $i$-th reaction is a scattering process, such as $\nu + N \rightarrow \nu + N$, on the other hand, we calculate the angles and energy after the scattering, $\theta^\prime_\nu, \phi^\prime_\nu$ and $E^\prime_\nu$, with random numbers mentioned in Section~\ref{new_MC}

\subsubsection{Emission} 
The total number of neutrinos emitted during a time step $dt_{\rm{f}}$ in unit spatial volume is calculated by the reaction rate $R_{i,\rm{ems}}$ and we add the corresponding number of sample particles uniformly in that volume element at the beginning of each time step.
The energies and angles of sample particles are distributed following the distribution of the reaction rate.
We put the distribution time into sample particles randomly in the range of [0, $dt_{\rm{f}}$] in order to get the constant emission rate and calculate their evolutions in the same way as those for existing sample particles.

%%%%%%%%%%%%%%%%%%%%%%%%%%%%%%%%%%%%%%%%%%%%%%%%%%%%%%%%%%%%%%%%%%%%%%%%%%%%%%%%%%%%%%%%%%%%%%%%%%%%%%%%%%%%%%%%%%%%%%%%%%%%%%%%%
\section{Neutrino reactions} \label{appendix3}

\subsection{Electron/positron scatterings}
The reaction rates of the electron/positron scattering are derived from the \textcolor{black}{similar form} as the nucleon scattering in eqs.~(\ref{R_rec})-(\ref{R_rec_f}), if we change the coefficients $\beta$'s summarized in Table~\ref{coef_esc}, the target mass $m_N \rightarrow m_e$ and the chemical potential $\mu_N \rightarrow \mu_e, -\mu_e$ for electrons and positrons, respectively.
In this paper, we denote the total reaction rates of electron and positron scatterings as $R_{\rm{esc}}$.
Their cross section $\sigma_{\rm{esc}}$ and normalized spectra $P_\psi$ and $P_{E^\prime_\nu}$ are defined in the same way as those for the nucleon scattering.
Note that we should distinguish the reaction rates of $\nu_x$ and $\bar{\nu}_x$, but we adopt that of $\nu_x$ in this study.

\begin{table*}[htbp]
\caption{The coefficients for the reaction rates of electron and positron scatterings. In this expression, $C^\prime_{Ve}= C_{Ve} + 1$ and $C^\prime_{Ae} = C_{Ae} + 1$ with $C_{Ve} = -1/2 + 2\sin^2{\theta_w}$ and $C_{Ae} = 1/2$. \label{coef_esc}}
\begin{center}
\begin{tabular}{c|ccc} \hline
reaction  & $\beta_1$ & $\beta_2$ & $\beta_3$ \\
\hline\hline
$\nu_e e^-$ / $\bar{\nu}_e e^+$ & $\left(C^\prime_{Ve} + C^\prime_{Ae}\right)^2$ &  $\left(C^\prime_{Ve} - C^\prime_{Ae}\right)^2$ &  $ C^{\prime2}_{Ae} - C^{\prime2}_{Ve} $ \\
$\nu_e e^+$ / $\bar{\nu}_e e^-$ & $\left(C^\prime_{Ve} - C^\prime_{Ae}\right)^2$ &  $\left(C^\prime_{Ve} + C^\prime_{Ae}\right)^2$ &  $ C^{\prime2}_{Ae} - C^{\prime2}_{Ve} $ \\
$\nu_x e^-$ & $\left(C_{Ve} + C_{Ae}\right)^2$ &  $\left(C_{Ve} - C_{Ae}\right)^2$ &  $ C^2_{Ae} - C^2_{Ve} $ \\
$\nu_x e^+$ & $\left(C_{Ve} - C_{Ae}\right)^2$ &  $\left(C_{Ve} + C_{Ae}\right)^2$ &  $ C^2_{Ae} - C^2_{Ve} $ \\
\end{tabular}
\end{center}
\end{table*}

\subsection{Electron capture on free proton and positron capture on free neutron}
The emission rate of EC's and PC's on free nucleons $R_{\rm{EC,ems}}$, $E_{\rm{PC,ems}}$ are calculated by \cite{1985ApJS...58..771B}:
\begin{eqnarray}
  R_{\rm{EC,ems}} &=& \frac{{G_F}^2}{\pi\hbar c}\eta_{\mathrm{pn}}\left({g_V}^2+3{g_A}^2\right)\left(E_{\nu_e}+Q\right)^2 \nonumber \\
 &&\times \sqrt{1-\frac{m_e^2}{\left(E_{\nu_e}+Q\right)^2}}f_e\left(E_{\nu_e}+Q\right), \\
  R_{\rm{PC,ems}} &=& \frac{{G_F}^2}{\pi\hbar c}\eta_{\mathrm{np}}\left({g_V}^2+3{g_A}^2\right)\left(E_{\nu_e}-Q\right)^2 \nonumber \\
  &&\times \sqrt{1-\frac{m_e^2}{\left(E_{\nu_e}-Q\right)^2}} f_{e^+}\left(E_{\nu_e}-Q\right) \nonumber\\
  &&\times\Theta\left(E_{\nu_e} - Q - m_e\right),
\end{eqnarray}
in which nucleons are non-relativistic and they neglect nucleon recoils.
The absorption rates are derived from the detailed balance relations, $R_{\rm{\ast,ems}} (1-f_{\ast,\rm{eq}}) = R_{\rm{\ast,abs}}f_{\ast,\rm{eq}}$, using the Fermi-Dirac distribution of electrons and positrons $f_{\ast,\rm{eq}}$ with the chemical potential $\mu_e$ for EC's and $-\mu_e$ for PC's; the reaction rates $R_{\ast,\rm{abs}} = R_{\rm{EC,abs}}, R_{\rm{PC,abs}}$.
The cross sections are calculated as $\sigma_{\ast} = R_{\ast,\rm{abs}}$.

%The expression of the emission rate for PC on free neutrons is very similar to that of EC on free protons except for the threshold for the neutrino energy \citep{1985ApJS...58..771B}.
%\begin{eqnarray}

%\end{eqnarray}

\subsection{Electron-positron pair annihilation}
We use the reaction rate of the electron-positron pair annihilation $R_{\rm{pair}}$\footnote{The reaction rate in \cite{2017ApJ...848...48K} is described in the natural unit ($c=\hbar=1$). In this paper, $R_{\rm{pair}}$ is defined by multiplying a factor $1/c\hbar$ to that in the previous paper.} described in \cite{2017ApJ...848...48K} (See eqs.~(1)-(9) in this paper).
The emission rate and cross section for neutrinos are derived from the integrals of the reaction rate in a phase space for anti-neutrinos:
\begin{eqnarray}
  R_{\rm{pair,ems}} &=& \int\int \frac{1}{2E_\nu \left(2\pi\right)^3} \frac{2\pi E^2_{\bar{\nu}}}{2E_{\bar{\nu}}\left(2\pi\right)^3} \nonumber \\
   &&\ \ \ \ \ \ \ \times R_{\rm{pair}} \left(1-f_{\bar{\nu}}\right) d\cos{\psi} dE_{\bar{\nu}}, \\
  \sigma_{\rm{pair}} &=&  \int\int \frac{1}{2E_\nu \left(2\pi\right)^3} \frac{2\pi E^2_{\bar{\nu}}}{2E_{\bar{\nu}}\left(2\pi\right)^3} \nonumber \\
  && \ \ \ \ \ \ \ \times R_{\rm{pair}} f_{\bar{\nu}} d\cos{\psi} dE_{\bar{\nu}},
\end{eqnarray}
with the energy of anti-neutrinos $E_{\bar{\nu}}$, the angle between 4 momenta of neutrino pair $\psi$ and the distribution for anti-neutrinos $f_{\bar{\nu}}$.
For anti-neutrinos, we integrate the reaction rate over $E_\nu$ instead of $E_{\bar{\nu}}$.
In this calculation, we employ the distribution function for the other neutrinos derived from the background CCSN simulations.

\subsection{Nucleon bremsstrahlung}
We calculate the reaction rate of the nucleon bremsstrahlung $R_{\rm{brem}}$ based on \cite{1979ApJ...232..541F,1987ApJ...316..691M}.
The emission and absorption rates $R_{\rm{brem,ems}}$, $R_{\rm{brem,abs}}$ and the cross section $\sigma_{\rm{brem}}$ are derived in the same way as those for pair annihilations.

%%%%%%%%%%%%%%%%%%%%%%%%

\end{document}